\documentclass[useAMS,usenatbib]{mn2e}

\usepackage{psfig, epsf, epsfig}

\title[Multiple stellar populations]{Secondary star formation 
within  massive star clusters: Origin of 
multiple stellar populations in globular clusters}
\author[K. Bekki]
{Kenji Bekki${}^1$\thanks{E-mail:
bekki@cyllene.uwa.edu.au} \\
${}^1$ICRAR M468
The University of Western Australia
35 Stirling Hwy, Crawley
Western Australia, 6009}

\begin{document}

\date{Accepted, Received 2005 February 20; in original form }

\pagerange{\pageref{firstpage}--\pageref{lastpage}} \pubyear{2005}

\maketitle

\label{firstpage}

\begin{abstract}

We numerically investigate whether and how gaseous ejecta from AGB stars
can be converted into new stars within originally massive star clusters (MSCs) 
in order to understand the origin of multiple stellar populations 
in globular clusters (GCs). 
We adopt a scenario in which (i) MSCs with masses of $M_{\rm s}$
can be formed from high-mass, high-density giant molecular
clouds (GMCs) in  their host galactic building
blocks embedded in dark matter halos at high redshifts
and (ii) their evolution therefore can be significantly influenced by  
$M_{\rm s}$,
their initial locations, and
physical properties of their hosts.
Our 3D hydrodynamical simulations
show that  gaseous ejecta from AGB stars can be retained
within MSCs and consequently  converted into
new stars
very efficiently in the central regions of MSCs, 
only if $M_{\rm s}$ exceed a threshold mass ($M_{\rm th}$)
of  $\approx 10^6 {\rm M}_{\odot}$.
The new stars can correspond to  the ``second generation (SG)'' of stars
with higher Na and lower O abundances observed in GCs.
Star formation efficiencies during the formation of SG stars
within MSCs with $M_{\rm s} \ge M_{\rm th}$
can be  rather high ($0.3-0.9$)
so that very compact new clusters within original MSCs
can be formed.  
$M_{\rm s}$ should be as large as $10^6-10^7 {\rm M}_{\odot}$
to explain the observed large fraction of SG stars in the present ordinary
Galactic GCs, because new stars can consist of only $1-4$\% among all stars 
for the standard IMF.
Nuclear  MSCs  
are found to retain much more effectively the AGB ejecta and convert
more efficiently 
the gas into new stars owing to much deeper gravitational potential
of their hosts.
Capture and accretion of cold molecular gas (or small GMCs) 
by forming MSCs themselves
can be mechanisms for mixing (i.e., dilution) of AGB ejecta with cold
pristine  gas.
We suggest that both  $M_{\rm s}$ and their locations
within their hosts
can determine whether abundance spread can be seen only in light elements
or even in heavy ones.
We discuss how and in what time scale MSCs preferentially
lose old stars
owing to tidal stripping by their host galactic building blocks.
We also suggest that the origin of the intermediate-age GCs with
possible age spread of $\sim 100$ Myr yet apparently no/little abundance spread 
in light elements in the LMC is closely associated with their incapability
to retain the AGB ejecta owing to their low masses.
\end{abstract}


\section{Introduction}

It has long been discussed both observationally and theoretically
why some of the Galactic GCs show star-to-star inhomogeneity
among the light elements of stars and what physical mechanisms 
are responsible for the inhomogeneity  (e.g., Cottrell \& Da Costa 1981;
Sneden et al. 1992; 
Norris \& Da Costa 1995; Cannon et al. 1998; 
Gratton et al. 2004; D'Antona \& Caloi 2004; Fenner et al. 2004;
Norris 2004; Lee et al. 2005; Smith et al. 2005;
Bekki et al. 2007; Alves-Brito et al. 2008; 
Calelan 2008; Kayser et al. 2008; Piotto 2008;
Da Costa et al. 2009; Marcolini et al. 2009; Yong et al. 2009;
Carretta et al. 2010; D'Ercole et al. 2010; 
Romano et al. 2010; van Loon 2010).
Although such star-to-star inhomogeneity was discovered
in 70's and  80's 
(e.g., Cohen 1978; Peterson 1980; Norris et al. 1981; Leep et al. 1986),
recent statistical studies for a larger number of the Galactic GCs
have established that the presence of multiple stellar populations 
is an universal phenomena seen across most of the Galactic GCs
(e.g., Carretta et al. 2010). 
The latest observational results by Ferraro et al. (2009)
have revealed that the Galactic  metal-rich GC Tarzan
5 has two different stellar populations with
different abundances of heavy elements. Lee et al. (2009) has also suggested
a significant fraction of the Galactic GCs have two different populations
with different abundances in heavy elements based on color-magnitude diagrams
of the GC stars
in $hk$-bands.

Now large star-to-star abundance variations have been
confirmed in almost unevolved stars in some of the Galactic GCs
(e.g., Gratton et al 2001;  Ramirez \& Cohen 2002; Bedin et al. 2004;
Gratton 2004;
Carretta et al. 2004; Piotto et al. 2005, 2007; D'Orazi et al. 2010a), which 
strongly suggests that after the first generation (FG) of stars
formed within forming GCs,  gas chemically mixed with gaseous ejecta
of  FG stars was converted into 
to  the second generation (SG) of stars (and even third and fourth
generations).
Observations have now being extensively investigating
how chemical properties of SG stars
(e.g. fractions of SG stars)  correlate with
their global internal properties (e.g., magnitudes and ellipticities)
and with locations and 3D motions
with respect to the Galactic center 
to understand the origin of SG stars and physical relationships
between formation of FG and SG stars (e.g., Carretta 2006, 
Carretta et al. 2010).

Following these observational developments,
theoretical studies have considered 
that SG stars can form from gaseous ejecta either by
AGB stars (``AGB scenario''; D'Antona \& Caloi 2004; Karakas et al. 2006)
or by fast rotating massive stars (``FRMS'' scenario; 
Prantzos \& Charbonnel 2006; Decressin et al. 2007)
and thereby investigated whether the observed physical
properties 
of the Galactic GCs can be explained by their models.
Although  it has not been determined  which of the two can explain 
observations in a self-consistent manner,
a number of authors suggested that
the AGB scenario 
is a more physically viable because  
AGB  ejecta 
is more likely to be  converted into new stars owing to low ejection
velocities
(e.g., Renzini 2008).

Recent observations have provided  the following two key results
which can give  strong constraints  
on any theory for GC formation: (i) the large fraction
(typically $0.67$) of SG stars in each individual Galactic GCs 
and (ii) the observed Na-O and Mg-Al
anti-correlations between cluster stars (e.g., Carreta et al. 2010
for a comprehensive study for these two).
The first key result suggests that the original stellar systems
(i.e., FG stars)
need to be much  more massive than the present GCs 
unless we adopt very unusual and unrealistic IMFs
(e.g., Smith \& Norris 1982; D'Antona \&  Caloi 2004):
the original systems are either very massive star clusters
or dwarf galaxies hosting GCs  (e.g., Bekki \& Norris 2006).
If the original systems are really massive ones, then the FG stars
need to be preferentially lost while SG ones remain  the same
so that the observed typical fraction ($\sim 0.67$)
of SG stars can be explained.

Although a number of theoretical works based on the AGB scenario 
tried to explain the second key observational result
in a self-consistent manner
(e.g., Fenner et al. 2004; Bekki et al. 2007; 
D'Antona et al. 2005; D'Antona \& Ventura 2007; 
Ventura \& D'Antona 2006, 2008;
D'Ercole et al. 2010),
their models appear to have not yet explained  all of the relevant
observations on chemical abundances of the Galactic GCs
in a fully self-consistent manner.
Chemical evolution models based on the FRMS scenario have not yet been fully
explored so that the validity of the FRMS scenario can not be currently
assessed.

In these previous models, it is assumed that
AGB ejecta can be converted into new stars (i.e., formation of SG stars)
within already existing clusters.
However it would not be so obvious that such secondary star formation
can occur within clusters, given the shallow gravitational potential
wells of clusters and the  small mass fraction of AGB ejecta. 
Therefore, 
secondary star formation processes within clusters 
should be investigated by numerical simulations that 
can include various physical processes within clusters
(e.g., retention of AGB ejecta).
D'Ercole et al. (2008) first investigated whether SG stars
can be formed from the gaseous ejecta of AGB stars of
FG with $M_{\rm s}=10^7 {\rm M}_{\odot}$.
However, their models are based on one-dimensional hydrodynamical
simulations and have  limitations in predicting
3D structures and kinematics of final stellar systems.
Three-dimensional (3D) stellar  and gas dynamical
numerical simulations with a plausible model for star formation
are  ideal  to investigate  secondary star formation within clusters
and can furthermore provide theoretical predictions that can be compared
with the observed differences in 3D structures and kinematics 
between FG and SG stars
(e.g.,  Norris et al. 1997; 
 Ferraro et al. 2002;
Sollima et al. 2005, 2007; Pancino et al. 2007
Bellini et al. 2009;
Anderson \& van der Marel 2010).

The purpose of this paper is thus 
to investigate extensively star formation from gaseous ejecta from AGB  stars
within MSCs based on self-consistent hydrodynamical simulations with
a reasonable model for star formation.
We adopt a scenario in which  (i) MSCs
can be formed 
in  their host galactic building
blocks embedded in dark matter halos at high redshifts
and (ii) new stars formed from AGB ejecta can finally become
the observed SG stars in the present GCs.
Based on this scenario,
we investigate (i) how AGB ejecta can be retained within MSCs,
(ii) whether and how secondary star formation
from AGB ejecta proceeds within MSCs, and (iii) how MSCs evolve if they
are located in nuclei of their hosts.
We also investigate how MSCs lose their stars during tidal interaction
with their hosts in order to discuss the observed smaller fractions 
of FG stars in the Galactic GCs.

A number of previous works discussed a scenario 
in which  GCs were originally 
stellar galactic nuclei
or nuclear star clusters in nucleated galaxies 
and their host galaxies had been already destroyed
by strong tidal fields of much larger galaxies to disappear completely 
(e.g., Zinnecker et al. 1988;  Freeman 1993; 
Bekki \& Freeman 2003; Bellazzini et al. 2008; B\"oker 2008).
Following this scenario,
Bekki et al. (2007) investigated chemical abundances of FG and SG stars
in  GCs
formed within  the central regions of  their host galaxies.
The evolution of MSCs initially nuclear regions of their hosts
in the present study therefore can provide some implications 
on the validity of the above scenario in explaining observational
properties of GCs.

The plan of the paper is as follows: In the next section,
we describe the details of the proposed scenario  for GC formation. 
In \S 3, we describe the numerical models for evolution
of gaseous ejecta of AGB stars within MSCs. 
In \S 4, we
present the numerical results
mainly on physical properties of SG stars.
In \S 5, we discuss how long MSCs can retain most of their stars
when they are influenced by strong tidal fields of their hosts. 
In \S 6, we discuss a number of key issues related to the origin
of multiple stellar populations in GCs.  
We summarize our  conclusions in \S 7.

\begin{figure}
\psfig{file=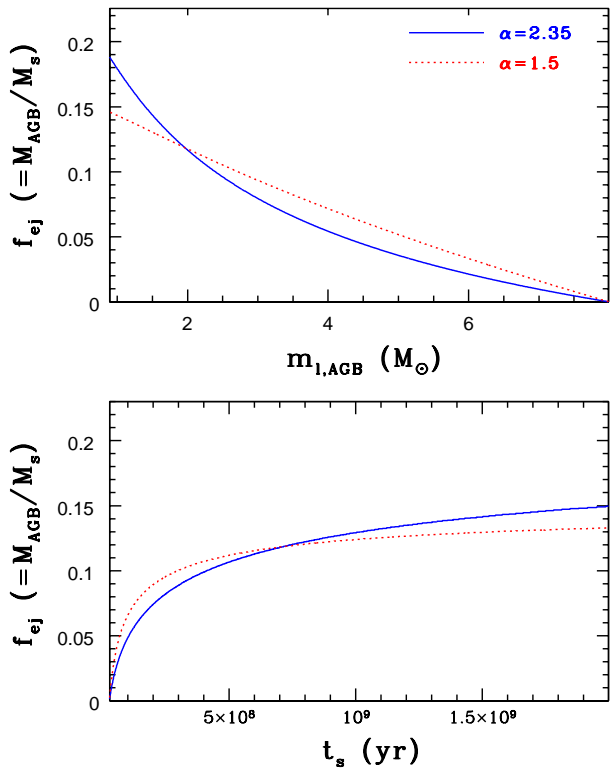,width=8.0cm}
\caption{
Dependences of the ratio ($f_{\rm ej}$)
of the total mass of AGB ejecta ($M_{\rm AGB}$)
to the total 
stellar mass of a MSC ($M_{\rm s}$) as a function
of the lower-mass cut-off of AGB stars ($m_{\rm l,AGB}$)
and the time that has elapsed since the most massive AGB 
star ($m_{\rm u, AGB}=8 {\rm M}_{\odot}$)
starts to eject gas for two different models with the IMF
slopes of $\alpha=2.35$ (blue solid) and 1.5 (red dotted).
This $f_{\rm ej}$ describes a possible maximum mass fraction of AGB ejecta
that can be converted into new stars within a MSC. 
Note that at most $\sim 10$\% of $M_{\rm s}$ can be possibly
converted into new stars within 1 Gyr.
}
\label{Figure. 1}
\end{figure}

\section{The scenario}

Although we  describe the scenario in the context of
the Galaxy formation,
the formation processes of GCs within  galaxies in general
would be similar to those described below.
The scenario is
based on the results from  our  present and previous theoretical studies
on GC formation
(e.g., Bekki et al. 2002; Bekki et al. 2004; Bekki \& Chiba 2007;  Bekki 2006;
Bekki et al. 2007; Hurley \& Bekki  2008;  Bekki et al. 2008).
Since the scenario is based partly  on
results of cosmological simulations (e.g., Bekki et al. 2008),
the scenario has some implications
on the observed correlations between abundance properties of GCs
and 3D motion and kinematics of GCs (e.g.,
Carretta 2006;  Lee et al. 2007).
In the present paper,  the masses of the present GCs
are represented by $M_{\rm gc}$ so that $M_{\rm s}$ (original
masses of the GCs) and $M_{\rm gc}$ can be discriminated with
each other.

\begin{table*}
\centering
\begin{minipage}{175mm}
\caption{Range of model parameters for numerical simulations of 
star-forming MSCs}
\begin{tabular}{cccccccc}
{$M_{\rm s}$
\footnote{The initial mass of a MSC in units of ${\rm M}_{\odot}$.}}
& {$R_{\rm s}$
\footnote{The initial size of a MSC in units of pc.}}
& {$s_{\rm rot}$
\footnote{The ratio of rotational energy to total kinetic one in
a MSC.}}
& {$f_{\rm AGB}$
\footnote{The ratio of the total mass
of AGB stars to that of stars in a MSC.}} 
& {Star formation
\footnote{``YES'' (``NO'') means that the star formation
model is (is not) included in the simulation. }}
& { ${\rho}_{\rm th}$
\footnote{The threshold gas density for star formation in units
of atoms cm$^{-3}$}}
& {External tidal field
\footnote{``YES'' (``NO'') means that a MSC can (can not) be influenced
by the tidal field of its host galaxy.}} 
& {$M_{\rm h}$
\footnote{The total mass of the host galaxy for a MSC 
in units of ${\rm M}_{\odot}$. }}\\
$10^4-10^8$  & $35-200$  & $0-0.32$ & $0.002-0.08$ 
& YES/NO  & $0-10^4$  & YES/NO & $10^8-2\times10^{10}$ \\
\end{tabular}
\end{minipage}
\end{table*}

\subsection{Time sequence}

\subsubsection{Formation of FG stars of MSCs within host dwarfs}

FG stars  of MSCs
form from high-density GMCs with masses ($M_{GMC}$) as large
as or larger than $10^7 {\rm M}_{\odot}$ within 
the Galactic building blocks which are later destroyed 
owing to the strong tidal field of the Galaxy
when they merge with the Galaxy. 
Most of the hosts form as massive dwarf galaxies with masses ($M_{\rm h}$)
larger than $10^9 {\rm M}_{\odot}$ before reionization ($z>10$),
and thus their star formation and chemical  enrichment histories 
can be different from those of the present dwarfs within the Galaxy
(Bekki et al. 2008).
The formation places of MSCs
are highly likely to their hosts' central regions where
mass fractions of GMCs can be much higher (i.e, not  necessarily
in the nuclei of the hosts).

The MSCs formed in the 
very centers of their hosts start their lives as nuclear MSCs
so that their evolution can be different significantly from
that for MSCs formed outside the very centers owing to much
deeper gravitational potentials of hosts' central regions.
The host GMCs of MSCs may well initially have numerous substructures
(smaller GMCs) owing to their large masses (e.g., Efremov 1995; 
Bonatto \& Bica 2010) so that MSCs at their birth can be composed of
a number of smaller clusters (i.e., not single entities).
These smaller sub-units can finally merge with one other to
form single massive clusters within the merging timescale depending 
on $M_{\rm s}$ and their sizes ($R_{\rm s}$).

\subsubsection{Evolution of FG stars within GMCs}

Massive stars and type II supernova can strongly influence 
later evolution of 
molecular gas left behind from the formation of FG stars.
If most of the energy from massive stars and type II supernova
are converted into kinetic energy of gas surrounding the stars,
then gaseous ejecta from the stars can not be mixed well with
the residual molecular gas (Bekki \& Chiba 2007).
Therefore,  SG formation from gaseous ejecta from the above energetic
stars and the mass fraction of SG stars formed from the mixed gas
is very small ($<10^{-4}$). However, if massive stars are FRMSs with
very slow ejection {\it radial} velocities
($<10$ km s$^{-1}$), then the gaseous ejecta  
could be well mixed with the residual gas to form SG stars before
supernova explosion expels a significant fraction of gas.

Although type II supernova can heat up and expel a significant fraction
of gas from massive GMCs,  some fraction of  gas can be
immune from the effects of
supernova owing to the clumpy nature of the massive GMCs. 
The smaller residual clouds may well be later accreted onto the 
forming clusters to dilute AGB ejecta. 
While supernova explosion is influencing significantly GMCs,
star formation rates within the GMCs can drop significantly.
Gaseous ejecta from massive AGB stars start to be accreted onto
the central regions of MSCs after all supernova events finish.
However star formation does not resume until an enough amount of gas
can be accumulated within the central regions: there should be
a threshold gas mass fraction ($f_{\rm ej,th}$) above which
star formation can start within MSCs. 
Therefore there can
be significant time-delay ($\sim 10^8$ yr)  between FG and SG formation in this scenario.

\subsubsection{Formation of SG stars}

SG stars in a MSC start to form efficiently in the central region of the MSC
when the mass fraction of the accumulated
AGB ejecta to $M_{\rm s}$ ($f_{\rm ej}$) exceeds $f_{\rm ej,th}$. 
This efficient secondary star formation 
{\it from AGB ejecta} can occur only if $M_{\rm s}$ exceeds
a threshold cluster mass ($M_{\rm th}$), because AGB ejecta can not
be efficiently retained in MSCs with smaller $M_{\rm s}$.
During this secondary star formation, the residual molecular gas 
can be mixed with AGB ejecta to form new stars. Owing to the
presence of deeper gravitational potential wells of MSCs,
star formation efficiencies (${\epsilon}_{\rm sf}$) can be 
much higher ($>0.3$) for this secondary star formation so that 
very compact clusters can be formed.
MSCs thus can initially show ``nested structures''
with more diffuse distributions of FG stars and more compact ones
of SG stars.

Although MSCs with $M_{\rm s}$ below $M_{\rm th}$ can hardly form new 
stars from AGB ejecta, such less massive clusters can still capture 
residual molecular gas and smaller nearby GMCs (not chemically contaminated
by ejecta from FG stars). Cold molecular gas 
with {\it initially very high-densities} obtained  by MSCs themselves
can be converted into new stars within the central regions
of MSCs (Bekki \& Mackey 2009).  Very low-mass MSCs  can be destroyed by
interaction with GMCs so that secondary star formation can not happen
in their  central regions. Therefore, there should be a threshold
cluster mass below which MSCs show differences neither in ages nor in
light elements among their stellar populations.
Secondary star formation processes
 within nuclear MSCs can be significantly different
from those described above owing to much deeper potential wells
of their hosts (as described in the present study).

\subsubsection{Tidal stripping of MSCs due to their host dwarfs}

Secondary star formation can continue until MSCs lose most of their FG stars,
though massive stars and type II supernova of SG stars 
can also suppress or even truncate further star formation.
Strong tidal fields  of hosts can efficiently strip stars preferentially
from FG stars
owing to initially diffuse spatial distributions.
This preferential stripping {\it by the Galaxy} has been already
proposed by D'Ercole et al. (2008) and discussed in the context
of origin of the Galactic stellar halo (Vesperini et al. 2010).
The present scenario suggests that  not the tidal field
of the Galaxy but those of hosts of MSCs are 
responsible for the tidal stripping of  most FG stars in MSCs.
The timescale of a MSC to lose most of their FG stars due to
tidal stripping by its host depends on its position,
with respect to its host ($R_{\rm p}$), $M_{\rm s}$, and $R_{\rm s}$.
Therefore, more massive and denser MSCs can continue secondary
star formation longer so that AGB stars with lower masses can
possibly contribute to further star formation.

\subsubsection{Evolution into the Galactic halo GCs}

Hosts of MSCs are strongly influenced by the tidal field of the forming
Galaxy during their merging with the Galaxy so that
they can be completely destroyed by the Galaxy.
The stripped stars from hosts, which include FG stars from MSCs,
can be some parts of the Galactic stellar halo.
MSCs are stripped during disintegration of their hosts to finally become
halo GCs. Strong ram pressure of the Galactic halo gas
(e.g., Frank \& Gisler 1976; Bekki 2006)  and no cold
gas available in the halo prevent almost completely star formation
of the MSCs since they become the Galactic halo GCs. 
Thus, multiple  stellar populations can be formed within MSCs only when
they are within their hosts.

Gas can be efficiently  transferred to the nuclear regions
of hosts  to be converted
into new stars while they are being destroyed by the Galaxy
(e.g., Bekki \& Freeman 2003).
These new stars can have chemical abundances
in heavy elements   different from those of
original nuclear MSCs, because they are from the outer regions, where
chemical enrichment histories are quite different from those of nuclei.
Nuclear MSCs can show star-to-star abundance variations not only
in light elements but also in heavy ones.

\subsection{Constraints on original masses of MSCs}

A number of authors have already suggested that if the standard
IMF is applied for FG stars and 
if ${\epsilon}_{\rm sf}$ (star formation efficiency for SG formation) 
is 1.0, then original total masses of FG stars ($M_{\rm s}$)
are required to be  at least ten times larger than the present masses
of SG stars in the Galactic GCs with multiple stellar populations
(e.g., Bekki \& Norris 2006).  Given that ${\epsilon}_{\rm sf}$ is not like
100\% as observed in star-forming regions in the Galaxy,
the required $M_{\rm s}$ should be even higher than the above. 
Recent chemical evolution models
have shown that the total mass of pristine gas that can mix with
AGB ejecta to form SG stars needs to be comparable to that of AGB ejecta
in order to explain the observed levels of star-to-star abundance
variations and the O-Na anti-correlation in the Galactic GCs  
(e.g.,  D'Ercole et al. 2010):
AGB ejecta should not be too much diluted by external gas captured/accreted
by MSCs. Thus the required $M_{\rm s}$ estimated in previous studies can not
change significantly even if contribution from external pristine
gas to SG formation is considered.

In the present scenario,  
gaseous ejecta from AGB stars with masses ranging from $m_{\rm l, AGB}$
to $m_{\rm u, AGB}$ can contribute to the formation of SG stars
within a  timescale of $t_{\rm s}$ (which determines $m_{\rm l,AGB}$).
We here estimate the mass fraction
of AGB ejecta to the initial mass of a MSC ($f_{\rm ej}$)
(i) for a given $m_{\rm l, AGB}$
and (ii) for a given $t_{\rm s}$
by assuming $m_{\rm u, AGB}=8 {\rm M}_{\odot}$ and power-law initial
mass functions with the slopes of $\alpha$. 
The adopted IMF in number defined
as $\psi (m_{\rm I}) = M_{s,0}{m_{\rm I}}^{-\alpha}$,
where $m_{\rm I}$ is the initial mass of
each individual star and the slope $\alpha =2.35$
 corresponds to the Salpeter IMF.
The normalization factor $M_{s,0}$ is a function of $M_{\rm s}$,
$m_{\rm l}$ (lower mass cut-off), and $m_{\rm u}$ (upper mass cut-off):
\begin{equation}
M_{s,0}=\frac{M_{\rm s} 
\times (2-\alpha)}{{m_{\rm u}}^{2-\alpha}-{m_{\rm l}}^{2-\alpha}}.
\end{equation}
where $m_{\rm l}$ and $m_{\rm u}$ are  set to be   $0.1 {\rm M}_{\odot}$
and  $120 {\rm M}_{\odot}$, respectively.
The total mass of AGB ejecta within a MSC ($M_{\rm AGB}$)
 is accordingly described as:
\begin{equation}
M_{\rm AGB}=\int_{m_{\rm l,AGB}}^{m_{\rm u,AGB}} m_{\rm ej} 
\psi (m) dm,
\end{equation}
where $m_{\rm ej}$ describes the total gas mass ejected from  
an  AGB star with initial mass  $m_{\rm I}$ 
and final mass  ($m_{\rm F}$).
We derive an analytic form of $m_{\rm ej}$
($=m_{\rm I}-m_{\rm F}$) from  the observational data
by Weidemann (2000) by using the least-square fitting method, and  
find:
\begin{equation}
m_{\rm ej} =0.916 M_{\rm I}-0.444.
\end{equation}

In order to calculate the main-sequence
turn-off mass ($m_{\rm TO}$)
we use the following formula
(Renzini \& Buzzoni 1986):
\begin{equation} 
\log m_{\rm TO}(t_{\rm s})
= 0.0558 (\log t_{\rm s})^2 - 1.338 \log t_{\rm s} + 7.764,
\end{equation} 
where $m_{\rm TO}$  is in solar units and time $t_{\rm s}$ in years.
%In the present study, the time $t_{\rm s}=0$
%corresponds to the epoch  when
%the most massive AGB star (with $m_{\rm u, AGB}=8 {\rm M}_{\odot}$)
%starts to eject gas in the present models.

Fig. 1 shows that $f_{\rm ej}$ is larger for smaller $m_{\rm l,AGB}$
both for the standard IMF ($\alpha=2.35$) and the top-heavy one
($\alpha=1.5$). Although the top-heavy model shows larger $f_{\rm ej}$ 
than the standard one for $m_{\rm l,AGB}<2 {\rm M}_{\odot}$,
the derived $f_{\rm ej}$ is small: only $0.036$ for 
$m_{\rm l,AGB} =5 {\rm M}_{\odot}$
and 0.079
$m_{\rm l,AGB} =3 {\rm M}_{\odot}$.
Ventura \& D'Antona (2008) showed that the observed O-Na anti-correlations
can be well reproduced only if 
AGB stars with masses larger $5 {\rm M}_{\odot}$ can contribute
to secondary star formation within GCs.
Thus, it is highly likely that progenitor clusters for the present GCs. 
needs to be at least 
$\sim 25$ times more massive than the present GCs.

As shown in  Fig. 1, 
$f_{\rm ej}$ can be larger than 0.1 for $t_{\rm s}>1$ Gyr 
(or $m_{\rm l, AGB}<3 {\rm M}_{\odot}$) for the two IMF models.
It should be here stressed that MSCs with larger $t_{\rm s}$
can end up with a large age difference ($>1$ Gyr) between FG and SG stars.
Original MSCs can lose most of their FG stars due to tidal stripping
of their hosts within well less than $\sim 1$ Gyr, as shown later in
the present study.
Therefore gaseous ejecta only from massive AGB stars
can participate secondary star formation within MSCs.
It would be reasonable to consider that $m_{\rm l, AGB}$ is
$4-5 {\rm M}_{\odot}$ that corresponds to $t_{\rm s}\sim10^8$ yr.

\begin{figure}
\psfig{file=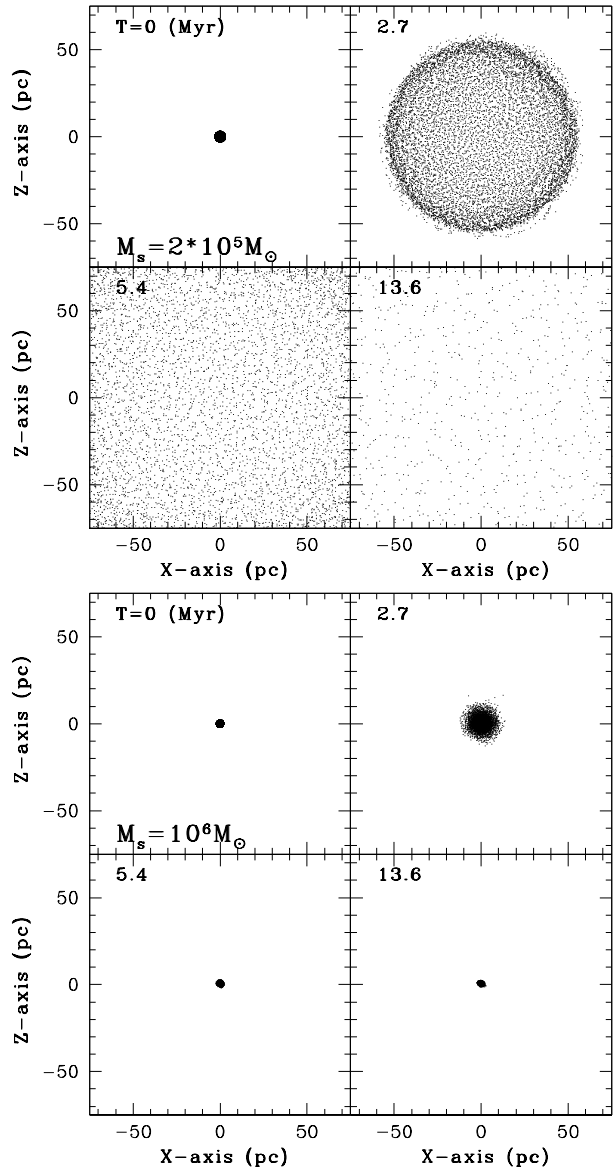,width=8.0cm}
\caption{
Time evolution of the distribution of gaseous ejecta from an AGB
star with $V_{\rm ej}=20$ km s$^{-1}$ projected onto the $x$-$z$ plane
for models with $M_{\rm s}=2\times 10^5 {\rm M}_{\odot}$ (upper four panels)
and $M_{\rm s}= 10^6 {\rm M}_{\odot}$ (lower four).
}
\label{Figure. 2}
\end{figure}

\begin{figure}
\psfig{file=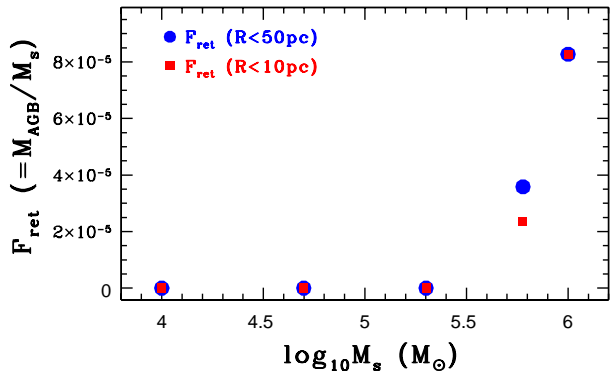,width=8.0cm}
\caption{
Dependences of the mass ratios ($F_{\rm ret}=M_{\rm AGB}/M_{\rm s}$)
of AGB ejecta that can finally be within the central 50pc (blue filled
circles) and 10pc (red filled squares) of MSCs
on $M_{\rm s}$.
}
\label{Figure. 3}
\end{figure}

\section{The numerical model}

We investigate star formation of
gaseous ejecta from FG AGB stars in MSCs with 
$M_{\rm s}$ ranging from
$10^4 {\rm M}_{\odot}$
to $10^8 {\rm M}_{\odot}$ by using
the latest version of GRAPE
(GRavity PipE, GRAPE-7) which is the special-purpose
computer for gravitational dynamics (Sugimoto et al. 1990).
We have revised our original GRAPE-SPH code (Bekki 2009)
so that we can investigate
star formation processes within the above-mentioned
massive stellar system.
MSCs with very large ($M_{\rm s} > 10^7 {\rm M}_{\odot}$) 
are still referred to as
``clusters'' just for convenience in the present study,
though their stellar masses
are as massive as those of  dwarf galaxies

We focus mainly on {\it secondary star formation} from AGB ejecta
of FG stars within original MSCs and describe the results of numerical
simulations on the star formation.
We have numerically  investigated
ram pressure stripping of AGB ejecta by ISM of their hosts
and its strength dependent on $M_{\rm s}$ and physical parameters of ISM.
We however describe the results
in our forthcoming papers (Bekki et al. 2010), 
because ram pressure stripping of AGB ejecta
of MSCs by typical ISM of hosts  is only effective
for low-mass MSCs with $M_{\rm s} < 10^4 {\rm M}_{\odot}$,
(which can not become ordinary GCs in the present study)
and thus not so important.

We have numerically  investigated
accretion of warm ISM onto MSCs by Bondi accretion
and found that
Bondi accretion with
the accretion rate of $\sim 10^{-3} {\rm M}_{\odot}$ yr$^{-1}$
is possible  only if $M_{\rm s}$ is larger than 
$3 \times 10^6 {\rm M}_{\odot}$ 
for typical ISM with $n=1$ atom cm$^{-3}$,  relative velocity
of $\sim 20$ km s$^{-1}$,  and gas temperature of 10000 K.
Although this accretion rate is too small for MSCs to obtain fresh gas
for the formation of SG stars  
within $\sim 10^8$ yr, this Bondi accretion can be an important
mechanism for obtaining fresh gas in 
massive GCs with abundance variations in heavy elements
like $\omega$ Cen, as briefly discussed later in this paper.
We  will describe the details of the numerical results
on Bondi accretion of ISM in 
our forthcoming papers (Bekki et al. 2010),
mainly because this paper becomes too long
if the results are included in this paper.

\subsection{Initial structures and kinematics of MSCs}

The original MSCs  are assumed to have a Plummer density profile
(e.g., Binney \& Tremaine 1987) with luminosities ($L_{\rm sc}$) and central
velocity dispersions (${\sigma}_{\rm s}$) consistent with the relation observed 
for GCs (Djorgovski et al. 1997):
\begin{equation}
L_{\rm s} = K_0 {{\sigma}_{\rm s}}^{1.7}, \;
\end{equation}
where $K_0$ is a normalization factor for the relation.
The scale length ($a_{\rm s}$) of a MSC is determined by the formula
\begin{equation}
a_{\rm s} = GM_{\rm s}/6{{\sigma}_{\rm s}}^{2}, \;
\end{equation}
where G is the gravitational constant.
Since the-mass-to-light-ratio ($M_{\rm sc}/L_{\rm s}$) is
assumed to be constant for all SCs,
$a_{\rm s}$ and ${\sigma}_{\rm s}$ are determined
by the equations (1) and (2) for a given $M_{\rm sc}$.
The normalization factor $K_0$
 in equation (5) is determined such that 
a cluster with $M_{\rm s}=6\times10^5$ $M_{\odot}$ can have 
the size of $R_{\rm s}=50$pc ($R_{\rm s}=5a_{\rm s}$), 
and the central velocity dispersion of 7 km s$^{-1}$.

Ha\c segan et al. (2005) revealed that 
massive stellar systems with masses larger than $2 \times 10^6 {\rm M}_{\odot}$
appear to have scaling regions different from those of the present GCs. 
We accordingly  consider that original sizes of MSCs with 
$M_{\rm s} \ge 2 \times 10^6 {\rm M}_{\odot}$
can not be determined by the above equations (5) and  (6) and 
therefore investigate
models with different $R_{\rm s}$ for a given $M_{\rm s}$
for such MSCs.
It is also possible that original MSCs with 
$M_{\rm s} <  2 \times 10^6 {\rm M}_{\odot}$ have 
$R_{\rm s}$ smaller or larger than those described by
the above scaling relations
because MSCs composed only of FG stars
have ages of $\sim 10^8$ yr in the present study: an order of  $10^8$ yr would 
not be long enough to form
GCs on the present scaling-relation owing to dynamical relaxation processes.
We therefore investigate models with $R_{\rm s}$ smaller and larger
than the above equation (5) and (6)  predict for a given $M_{\rm s}$.

A MSC is assumed to have
a small amount of  rotation with
the ratio of the  initial rotational energy ($T_{\rm rot}$)
to the total kinetic one ($T_{\rm kin}$) being a free parameter represented
by $s_{\rm rot}$.
 The parameter values of $s_{\rm rot}$ range
from 0 (no rotation) to 0.3 (rapid rotation).
The initial rotational velocity of a particle at a distance of $R$ from
the center of MSC is $\omega R$, where $\omega$ (constant angular velocity)
is determined
such that $s_{\rm rot}$ can be the adopted value.
Therefore, the system has random kinetic energy ($T_{\rm ran}$)  of
$(1-s_{\rm rot})T_{\rm kin}$ (due to isotropic velocity dispersion
of stars).
Firstly we estimate $T_{\rm kin}$ for ${\sigma}_{\rm s}$ determined
by $M_{\rm s}$ and $R_{\rm s}$ (in equations (5) and (6)) and then reduce 
${\sigma}_{\rm s}$ 
so that the final system can be in virial equilibrium
($T_{\rm kin}=T_{\rm ran}+T_{\rm rot}=0.5 |W|$, where $W$ is the total
potential energy of the system).

\subsection{Gas dynamics}

The FG stars in a MSC are  represented by 
equal-mass 
stellar particles with the particle number of $N_{\rm s}$ ($=10^5$) 
and some fraction of the particles are assigned as ``AGB stars''
with initial masses of $m_{\rm s}$
and the total number of $N_{\rm AGB}$
that can eject SPH gas particles with ejection velocities
of $V_{\rm ej}$  with respect to the
centers of the AGB stars.
The mass  fraction of AGB particles in a MSC 
is a parameter represented
by $f_{\rm AGB}$ that is determined by the adopted IMF.
As outlined in \S 2,
gaseous ejecta only from massive AGB stars with
initial masses of $\sim 5-8 {\rm M}_{\odot}$ need to be
converted into new stars (i.e., SG stars) to explain
the chemical abundances of SG (e.g., 
Ventura \& D'Antona 2008).
In the present paper, $V_{\rm ej}$=20 km s$^{-1}$
is adopted, which is a reasonable choice
for massive AGB stars (e.g., Marshall et al. 2004).

The present simulations can not resolve gaseous evolution of each
individual AGB star owing to the adopted numerical resolution
($\sim 0.5$ pc). We therefore assume that each AGB particle initially
has an expanding gaseous sphere which is much larger than the AGB star itself.
The mass, size, and  temperature of the large gaseous sphere
represented by SPH particles with the particle number
of $n_{\rm AGB}$   are set to be
$m_{\rm g}$, $r_{\rm g}$ and $T_{\rm g}$, respectively.
Each AGB particle accordingly  has a gas sphere represented
by $n_{\rm AGB}$ SPH particles with radial velocities (with respect
to the AGB particle) of $V_{\rm ej}$.  The AGB particle therefore
has a mass of $m_{\rm s}-m_{\rm g}$ after gas ejection.
 We here assume that each AGB particle
in a model
eject SPH particles at $T=0$ when the simulation starts: we do not consider
gradual ejection of gas from AGB stars at each time step, because
an appropriate modeling of such gradual ejection
requires a huge number of gas particles and thus is numerically
costly (i.e., practically not feasible).

Although we have investigated models with $T_{\rm g}=100 - 1000$K,
we mainly describe  the results of the models with $T_{\rm g}=100$K
which corresponding to star-forming warm molecular clouds
(e.g., Wilson et al. 1997).
This is because we consider that AGB wind can cool down  during expansion
throughout interstellar space due to radiative cooling
(to finally become molecular gas for further star formation)
so that $T_{\rm g}$ can becomes
much smaller than the original temperature of the wind ($\sim$ 1000K).
We adopt  an isothermal equation 
so that AGB ejecta can keep initially low temperatures
($100$K).
It should be stressed here that if $T=1000$K is adopted
(which is not realistic though),
star formation is possible only  for  models
with $M_{\rm s} \ge  10^6 {\rm M}_{\odot}$.

Most of AGB ejecta can be converted into new stars
well before  a few  Myr after commencement of secondary star formation
so that feedback effects of massive stars and type II supernova
can not significantly influence 
gas dynamics during the  secondary star formation processes.
We therefore consider that the above isothermal assumption 
is reasonable. 
In order to estimate value of  $m_{\rm g}$
corresponding  to the total mass ejected from each
AGB star after the main-sequence turn-off, we use
the formula given in the equation (3). 
About 83\% of a AGB particle
with $m_{\rm s}=5 {\rm M}_{\odot}$ can be ejected to be
used for further star formation in the present model.

If we adopt the Salpeter IMF, 
then $f_{\rm AGB}$ 
(the mass fraction of AGB stars with masses  larger than $5 {\rm M}_{\odot}$)
is $\sim 0.04$ in
the present study.
Since initial  masses within a MSC are the same between stellar particles,
$N_{\rm AGB}=f_{\rm AGB}N_{\rm s}$.
Although we adopt this $f_{\rm AGB}=0.04$
for most models,  we investigate models with different
$f_{\rm AGB}$ so that we can find  a threshold $f_{\rm AGB}$
for a given MSC 
above which secondary star formation can occur in the MSC.
It is found that if $f_{\rm AGB}$ is larger than 0.008, secondary
star formation is possible in some models, though the star formation efficiency
is low. This means that secondary star formation within a MSC
can not be possible until a certain amount of gas is accumulated
within the MSC and thus that there should be a time-delay between
the formation of FG stars and that of SG ones.

\begin{figure}
\psfig{file=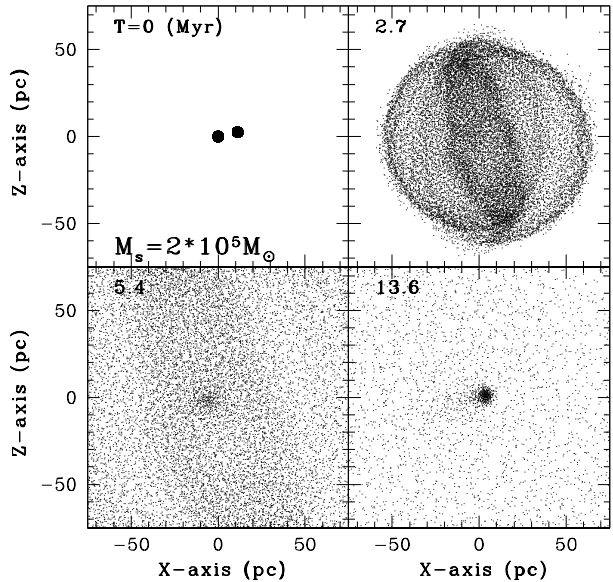,width=8.0cm}
\caption{
The same as Fig. 2 but for the model with two AGB stars in
a MSC with $M_{\rm s}=2\times 10^5 {\rm M}_{\odot}$.
}
\label{Figure. 4}
\end{figure}

\begin{figure}
\psfig{file=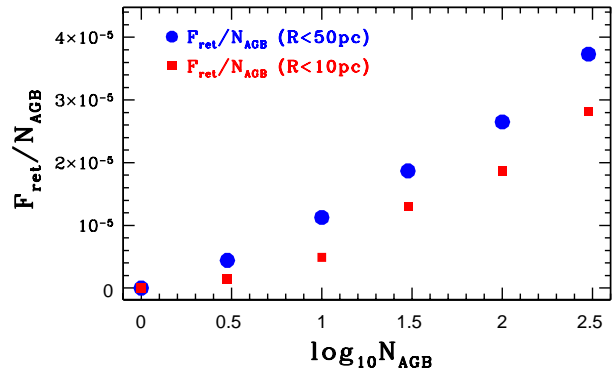,width=8.0cm}
\caption{
Dependences of $F_{\rm ret}$ normalized by $N_{\rm AGB}$ on
$N_{\rm AGB}$ 
in a  MSC with $M_{\rm s}=2\times 10^5 {\rm M}_{\odot}$.
This $F_{\rm ret}/N_{\rm AGB}$
describes how efficiently AGB ejecta  can be retained within MSCs.
}
\label{Figure. 5}
\end{figure}

\begin{figure}
\psfig{file=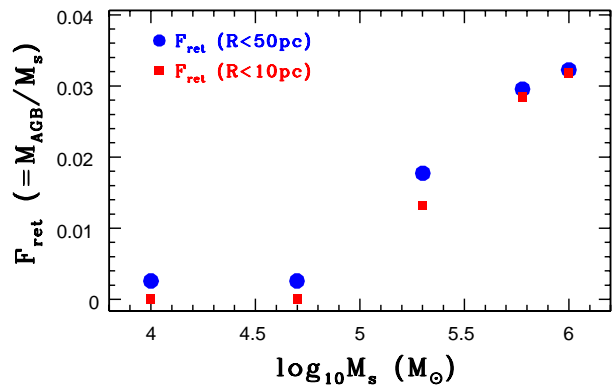,width=8.0cm}
\caption{
The same as Fig.3 but for the models in which all AGB stars can eject
gas with $V_{\rm ej}=20$ km s$^{-1}$.
}
\label{Figure. 6}
\end{figure}

\subsection{Star formation}

We investigate whether the gas accumulated
in the central regions of MSCs
can be sufficient to form new stars by adopting a
simple prescription for star formation.
In the models with ``star formation'',
a gas particle is converted into a collision-less new stellar
one if the gas particle meets the following three conditions:
(i) the dynamical time scale of the SPH gas particle
is shorter than the sound crossing time,  (ii) the gas is converging
(i.e., $\nabla {\bf v} <0$, where ${\bf v}$ is the velocity vector of the
gas particle),
and (iii) the local gas density exceeds a  threshold gas
density (${\rho}_{\rm th}$) which corresponds to the densities of 
dense cores of molecular clouds ($n > 10^4$ atoms cm$^{-3}$),
where individual star formation is ongoing.
The first  two conditions mimic the Jeans gravitational instability
for gaseous collapse. The new stellar particles
and old ones initially within MSCs  are referred to
as ``new stars'' (i.e., SG stars) 
and ``old stars'' (i.e., FG ones), respectively.
Also original clusters and new ones
formed from AGB ejecta are referred to as   ``old clusters''
and ``new clusters'', respectively, just for convenience.

We investigate models with ${\rho}_{\rm th}=0$  and $10^4$ 
atoms cm$^{-3}$ in order to investigate (i) whether
the threshold gas density is important for determining star formation histories 
of SG stars and (ii) how the present results,
in particular, structural and kinematical properties of final clusters
composed of SG stars,
depend on ${\rho}_{\rm th}$: we here consider that ${\rho}_{\rm th}$
has not been observationally well determined yet
and therefore could  be different for individual star forming clouds.
Although these star formation models are less realistic in some points
(not inclusive of magnetic fields and radiative transfer etc),
we consider that they enable us to grasp essential ingredients of
secondary star formation from AGB ejecta within original massive clusters.

\begin{figure*}
\psfig{file=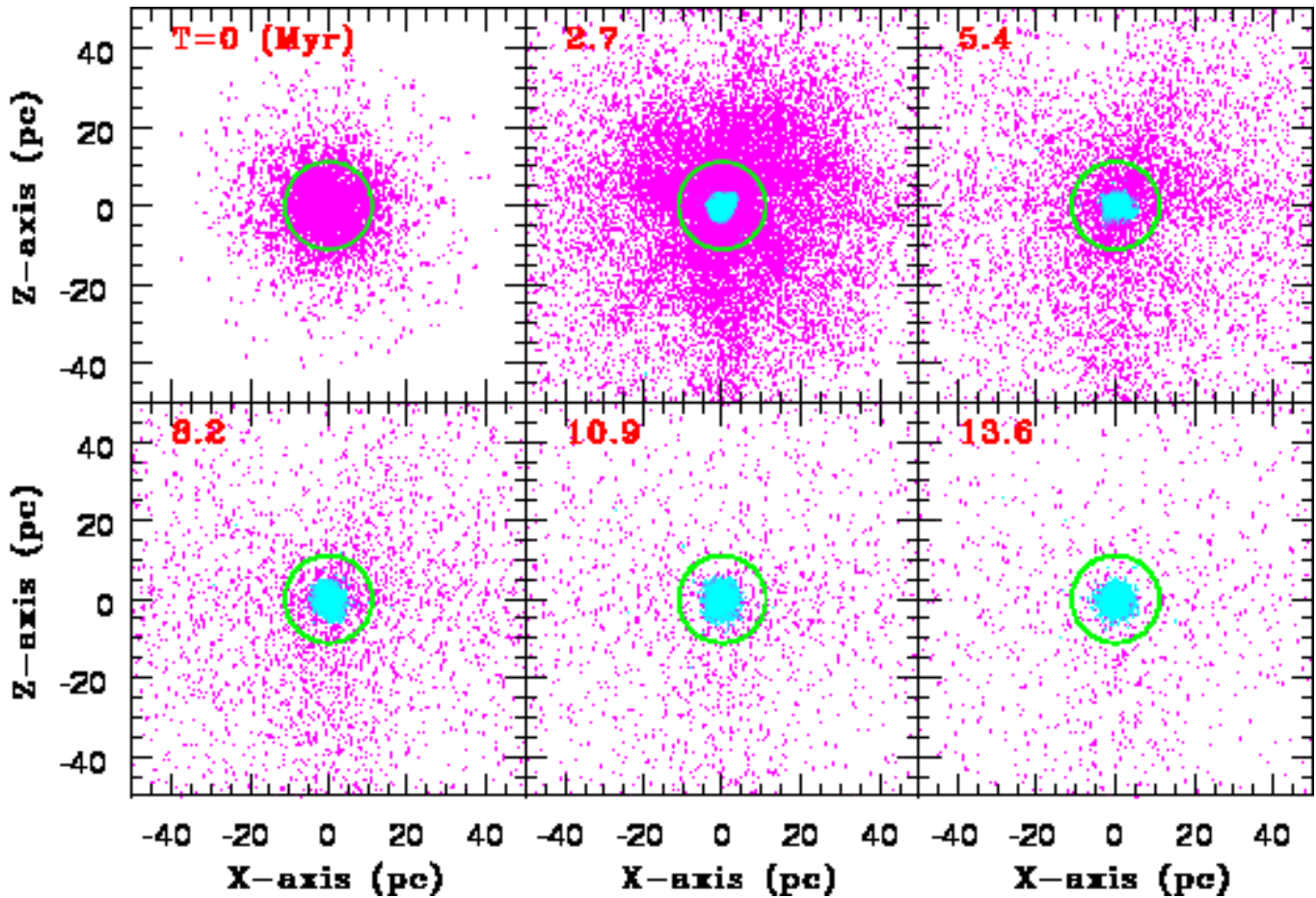,width=18.0cm}
\caption{
Time evolution of the distributions of gas (magenta) and new stars (cyan)
projected onto the $x$-$z$ plane for the standard model with
$M_{\rm s}=10^6 {\rm M}_{\odot}$,
$R_{\rm s}=44.7$pc,
$s_{\rm rot}=0.0$,
and
${\rho}_{\rm th}=10^4$ atoms cm$^{-3}$.
The green circle in each frame represent the original half-mass
radius of old stars. Time ($T$ in units of Myr)
that has elapsed since the simulation started
is shown in the upper left corner for each frame.
}
\label{Figure. 7}
\end{figure*}

\subsection{External gravitational  fields of host galaxies}

We mostly investigate ``isolated MSCs'' which are located outside
nuclei of their host galaxies and therefore
short-term ($\sim 10^7$ yr)
evolution of AGB eject can not be strongly influenced
by the tidal fields of their hosts. 
However, dynamical evolution of AGB ejecta from 
nuclear MSCs that are initially located in the very centers
of their hosts can be significantly influenced by their hosts
owing to the much deeper gravitational
potential wells of the hosts.
We therefore investigate how hosts influence evolution of AGB ejecta
from nuclear MSCs by assuming that the host dwarfs are dominated
by dark matter halos with masses of $M_{\rm h}$
and thus their gravitational  potentials are determined
only by the dark halos.

We adopt the density distribution of the NFW
halo (Navarro, Frenk \& White 1996) suggested from CDM simulations:
\begin{equation}
{\rho}(r)=\frac{\rho_{0}}{(r/r_{\rm s})(1+r/r_{\rm s})^2},
\end{equation}
where  $r$, $\rho_{0}$, and $r_{\rm s}$ are
the spherical radius,  the characteristic  density of a dark halo,  and the
scale
length of the halo, respectively.
Although we investigate models with 
$10^8 {\rm M}_{\odot} \le M_{\rm h} \le 2\times 10^{10} {\rm M}_{\odot}$,
we mainly describe the results of
the  models with $M_{\rm h}=10^9 {\rm M}_{\odot}$,
$r_{\rm s}=0.6$ kpc, and the virial radius ($r_{\rm vir}$) of 9.9 kpc
(i.e., the $c$-parameter of 13.9).
These models with $M_{\rm h}=10^9 {\rm M}_{\odot}$ are reasonable 
in the present scenario in which 
there is a threshold galaxy mass
($10^9 {\rm M}_{\odot}$)
above which GCs can be formed.
It should be here stressed that low-mass dark matter halos with 
$M_{\rm h} = 10^8 {\rm M}_{\odot}$ can  influence
evolution of AGB ejecta in hosts 
as significantly as  those with $M_{\rm h}=10^9 {\rm M}_{\odot}$.

\subsection{Ranges of parameters}

Firstly we investigate (i) whether AGB ejecta
can be retained within MSCs and (ii) whether and how
secondary star formation from AGB ejecta can proceed
for models that are consistent with the scaling relation
of GCs shown in the equation (2). 
Secondly we 
investigate structure and kinematics
of the simulated clusters in the models with $M_{\rm s}=10^7 {\rm M}_{\odot}$
and $R_{\rm s}=100$pc in which the final masses of SG stars can be 
as large as $\sim 10^5 {\rm M}_{\odot}$.
Thirdly we investigate whether much deeper gravitational potential
wells of host dwarf galaxies can play a role in  retaining
AGB ejecta efficiently in
the nuclear MSCs.

We have investigated secondary star formation processes
for numerous  models with different model parameters
(e.g., $M_{\rm s}$, $R_{\rm s}$, ${\rho}_{\rm th}$, and $s_{\rm rot}$);
we mainly describe here the results for the ``standard model''
with $M_{\rm s}=10^6 {\rm M}_{\odot}$, $R_{\rm s}=44.7$pc,
$s_{\rm rot}=0.0$, and ${\rho}_{\rm th}=10^4$ atoms cm$^{-3}$.
We assume that $n_{\rm AGB}=20$ for most models except those
for investigating how AGB ejecta can be retained within MSCs.
The reason for this adoption is that as long as $n_{\rm AGB}>10$,
the results do not depend on $n_{\rm AGB}$. 
Since we focus on secondary star formation processes with
a timescale of $<10^7$ yr,
we do not intend to discuss long-term dynamical evolution
of clusters due to two-body relaxation. We accordingly introduce
a gravitational softening length (${\epsilon}_{\rm g}$) 
for each simulation and 
${\epsilon}_{\rm g}$ is set to be equal to the mean particle
separation at the half-number radius of old stars.

The ranges of model parameters are shown in the Table 1.
We describe in detail only the results of some representative
models showing key parameter-dependences in retention processes 
of AGB ejecta and  secondary star formation
processes within clusters 
among the investigated models.
Secondary star formation processes can  continue only for $\sim 10^7$ yr
so that they can not be strongly influenced  
by  gravitational fields of their hosts unless 
MSCs are located in the very centers of their hosts.
We thus show mainly  the results for models without gravitational  fields
of hosts in order to much more 
clearly show the importance of the four key parameters $M_{\rm s}$,
$R_{\rm s}$, $s_{\rm rot}$, and ${\rho}_{\rm th}$ in secondary
star formation processes.

\subsection{Limiations of the model}

As described in \S 3.2,  we do not consider that different AGB stars with different
masses eject gas at different $T$ owing to different lifetimes of the stars. 
Therefore all of the gaseous ejecta can be converted into new stars 
within $\sim 10$ Myr. This rapid consumption of the AGB ejecta is not
so realistic, given that there should  be at least several tens millions years 
time delay between the epochs when AGB stars with masses of $8 {\rm M}_{\odot}$
and $5 {\rm M}_{\odot}$ start to eject
their gas. If we consider the different epochs of gas ejection from AGB stars
with different masses, then the star formation period would become significantly longer.
However,  energy feedback effects of 
numerous supernovae from the SG stars are likely to truncate star formation
in gaseous ejecta from lower-mass AGB stars.
In our future papers, we will discuss this point in detail
by using a more sophisticated numerical model.

The present model does not include the effect of type Ia supernova (SN Ia) on
star formation processes of SG stars. D'Ercole et al. (2008) already showed
that the cumulative effect of numerous SN Ia is important for the star formation
processes  
within massive clusters, because it can drastically alter gas dynamics within
the clusters.  Therefore  the present model that does not include
such SN Ia effect  possibly
overestimates  the star formation rates in the accumulated AGB ejecta in the
central regions of MSCs. 
Our future more sophisticated models with SN Ia effect will discuss how
SN Ia can influence gas dynamics and star formation in MSCs.

\begin{figure}
\psfig{file=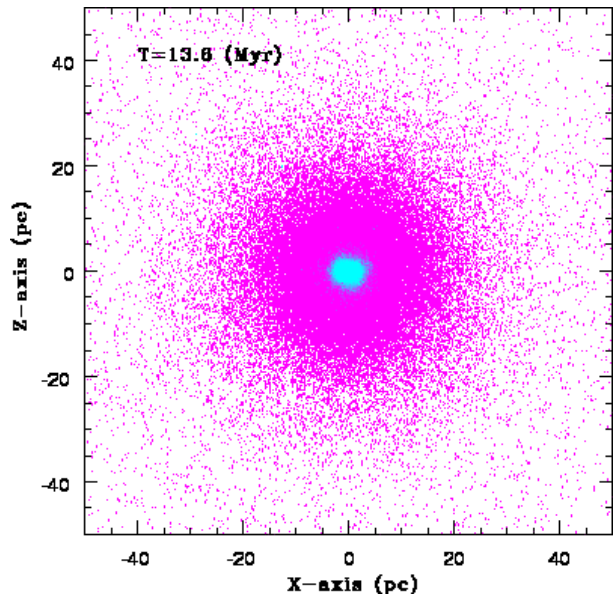,width=8.0cm}
\caption{
The final distributions of gas (magenta) and new stars (cyan)
projected onto the $x$-$z$ plane at $T=13.6$ Myr for the standard model.
}
\label{Figure. 8}
\end{figure}

\begin{figure}
\psfig{file=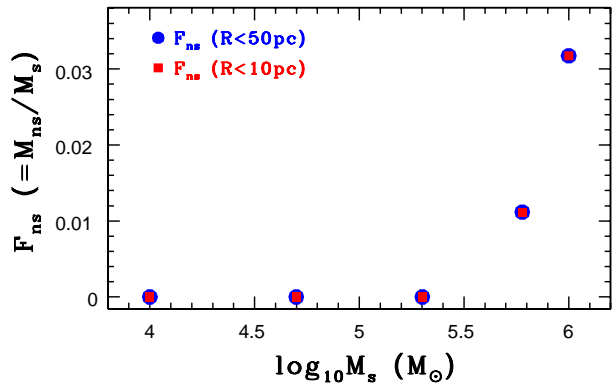,width=8.0cm}
\caption{
Dependences of mass fraction of new stars ($F_{\rm ns}=M_{\rm ns}/M_{\rm s}$)
on $M_{\rm s}$
for $R<50$pc 
(blue filled circles) and $R<10$pc (red filled squares). 
}
\label{Figure. 9}
\end{figure}

\begin{figure}
\psfig{file=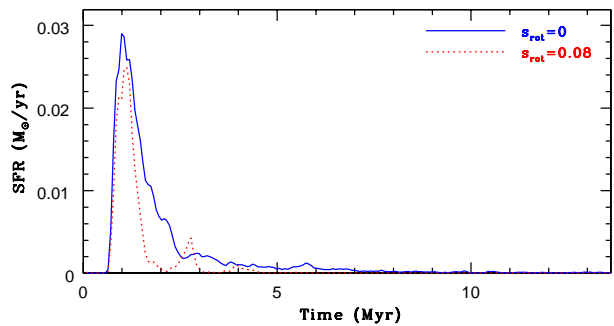,width=8.0cm}
\caption{
Star formation histories for the standard model
with $s_{\rm rot}=0$  (blue solid)
and the model with $s_{\rm rot}=0.08$ and other parameters being
the same as those of the standard model (red dotted). 
}
\label{Figure. 10}
\end{figure}

\section{Results}

\subsection{Retention of AGB ejecta}

Fig. 2 show that time evolution of spatial distributions of AGB ejecta
is very different between models with 
$M_{\rm s}=2\times 10^5 {\rm M}_{\odot}$ and
$M_{\rm s}= 10^6 {\rm M}_{\odot}$ in which model parameters 
except $M_{\rm s}$ are exactly the same. 
Gaseous particles with $n_{\rm AGB}=10^4$  ejected from an AGB star
locating in the very center of a MSC  can  expand quickly ($T=2.7$Myr) 
and finally  escape from the cluster and never
return back in the model with
$M_{\rm s}=2\times 10^5 {\rm M}_{\odot}$ ($T=13.6$Myr).
On the other hand,
gaseous ejecta from  an AGB star in the model
with $M_{\rm s}= 10^6 {\rm M}_{\odot}$
can return back to the initial location within a timescale of a few  Myr
after the ejection owing to the deeper gravitational potential
for this more massive cluster.
This clear difference suggests that $M_{\rm s}$
is an important parameter for determining  
whether AGB ejecta can be retained in MSCs during their early formation
phases.

Fig. 3 shows how the mass fractions
of AGB ejecta retained within MSCs ($=F_{\rm ret}$)
depend on $M_{\rm s}$.
Fig. 3 shows that MSCs with
$M_{\rm s} > 2\times 10^5 {\rm M}_{\odot}$ 
can retain a significant fraction of AGB ejecta if a  scaling-relation
similar to the observed one is applied for all MSCs.
This result implies that there can be a threshold mass above which
AGB ejecta can be efficiently retained.
For MSCs with 
$M_{\rm s} \sim  2\times 10^5 {\rm M}_{\odot}$,
AGB ejecta initially in the outer part of the MSC
can escape from the MSC whereas that initially in the inner part
can be retained within the MSC. As a result of this,
a ``core-halo'' structure of AGB ejecta can be formed within
13.6 Myr.
A very compact gas sphere can be formed where star formation 
would  be able to  proceed
rapidly if star formation is included in the model with 
$M_{\rm s} =10^6 {\rm M}_{\odot}$.
A larger fraction of AGB ejecta can be retained in MSCs with
larger $M_{\rm s}$.

Gaseous spheres composed of AGB ejecta can interact with one another
within a MSC when numerous stars enter into AGB phase almost simultaneously.
The present 3D numerical simulations enable us to investigate whether and
how hydrodynamical interaction between the gaseous spheres 
increase
or decrease the total gas mass accumulated within the central region
of the MSC.
Fig. 4 shows that if  two AGB stars
with each having $n_{\rm AGB}=10^4$  are included in 
the model with
$M_{\rm s} =  2\times 10^5 {\rm M}_{\odot}$
(where no gas accumulation is seen in Fig. 2),
the gas 
can lose kinetic energy owing to energy dissipation during
hydrodynamical interaction of the gaseous spheres 
so that  a non-negligible amount of gas can be accumulated within
the cluster.  About 3.6\% of the initial AGB ejecta can be accumulated
within the central 50pc of the cluster within 13.6 Myr.

Fig. 5 shows how $F_{\rm ret}/N_{\rm AGB}$,
where $N_{\rm AGB}$ are the total numbers of AGB stars,
depends on $N_{\rm AGB}$: this normalized
$F_{\rm ret}$ can measure the efficiency of retaining AGB ejecta within MSCs.
It is clear from this figure that
AGB ejecta can be more efficiently retained within clusters
for models with larger 
$N_{\rm AGB}$.
This is mainly because as the larger number of gaseous spheres (of AGB stars)
hydrodynamically interact with one another, the larger amount of
the gas can lose their kinetic energy to be accreted onto the central
regions of MSCs.
The results in Figs. 4 and 5 therefore demonstrate that 
hydrodynamical interaction of AGB ejecta can cause efficient
gaseous dissipation and thus play a great role in retaining
AGB ejecta within MSCs.

Fig. 6 shows that (i) there is a threshold $M_{\rm s}$
($=2 \times 10^5 {\rm M}_{\odot}$ in this parameter study)
above which AGB ejecta can be retained effectively
and (ii) $R_{\rm ret}$ is larger for larger $M_{\rm s}$.
The reason of these dependences is that MSCs with
larger $M_{\rm s}$ have deeper gravitational potentials
so that they can retain AGB ejecta more effectively.
Differences in $F_{\rm ret}$ between $R<10$pc and $R<50$pc
are small for models with $M_{\rm s}\ge 6 \times 10^5 {\rm M}_{\odot}$,
which means
that most the accumulated gas can form very compact gaseous regions 
within the central regions of MSCs.
Thus MSCs with larger $M_{\rm s}$ can obtain a larger amount of
AGB ejecta for further star formation: $M_{\rm s}$ is a key for determining
whether MSCs can have SG stars.

\begin{figure}
\psfig{file=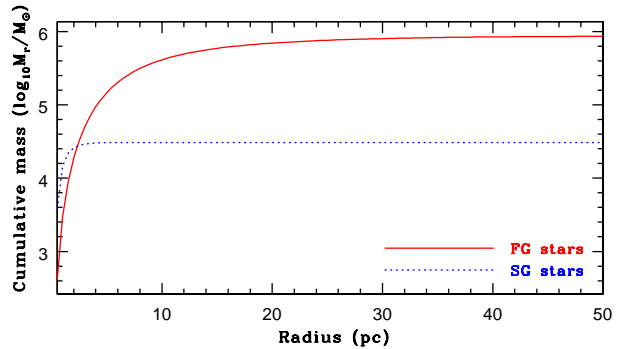,width=8.0cm}
\caption{
The cumulative mass within $R$ ($M_{\rm r}$) as a function of distance
($R$) from the center of a  MSC 
for FG stars (red solid) and SG ones (blue dotted) 
in the standard model. These FG and SG stars
are old and new ones, respectively, in the present study.
}
\label{Figure. 11}
\end{figure}

\begin{figure}
\psfig{file=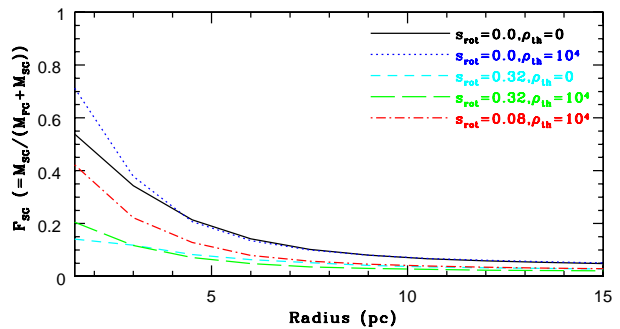,width=8.0cm}
\caption{
The mass fraction of SG stars 
($F_{\rm SG}=M_{\rm SG}/(M_{\rm FG}+M_{\rm SG})$)
as a function of distance from the center of a MSC  ($R$)
for different four models with
$s_{\rm rot}=0$ and ${\rho}_{\rm th}=0$ atoms cm$^{-3}$
(black solid),
$s_{\rm rot}=0$ and ${\rho}_{\rm th}=10^4$ atoms cm$^{-3}$
(blue dotted),
$s_{\rm rot}=0.32$ and ${\rho}_{\rm th}=0$ atoms cm$^{-3}$
(cyan short-dashed),
$s_{\rm rot}=0.32$ and ${\rho}_{\rm th}=10^4$ atoms cm$^{-3}$
(green long-dashed),
and $s_{\rm rot}=0.08$ and ${\rho}_{\rm th}=10^4$ atoms cm$^{-3}$
(red dot-dashed).
The initial masses of MSCs in these models are the same as that
of the standard model ($M_{\rm s}=10^6 {\rm M}_{\odot}$).
}
\label{Figure. 12}
\end{figure}

\begin{figure}
\psfig{file=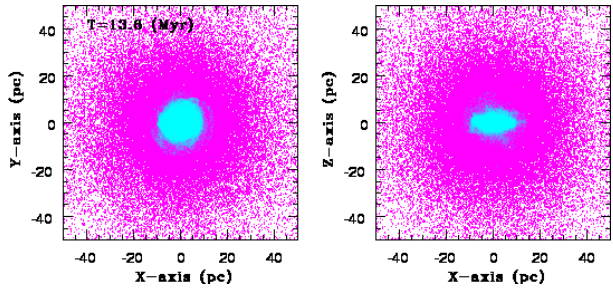,width=8.0cm}
\caption{
The final distributions of old stars (magenta) and new ones (cyan)
projected onto the $x$-$y$ plane (left) and the $x$-$z$ one (right)
in the more massive model with
$M_{\rm s}=10^7 {\rm M}_{\odot}$,
$R_{\rm s}=100$pc,
$s_{\rm rot}=0.02$,
and
${\rho}_{\rm th}=10^4$ atoms cm$^{-3}$.
}
\label{Figure. 13}
\end{figure}

\subsection{Secondary star formation}

Fig. 7 shows how gas ejected from AGB stars is accumulated within 
the central region of a MSC and consequently 
converted into new stars (i.e., SG ones) there within 
13.6 Myr in the  standard model with
$M_{\rm s}=10^6 {\rm M}_{\odot}$,
$R_{\rm s}=44.7$pc, 
$s_{\rm rot}=0.0$,
and 
${\rho}_{\rm th}=10^4$ atoms cm$^{-3}$.
The gaseous ejecta from AGB stars initially located in
the central region of the MSC can be  accumulated
there  within $\sim 1$Myr so that 
star formation can start efficiently.
The star formation from AGB ejecta can proceed like
a ``starburst'' with most gas being consumed rapidly within the first
3 Myr evolution of the MSC.
The star formation rate (SFR) 
reaches  its maximum value
($0.03 {\rm M}_{\odot}$ yr$^{-1}$) at $T=1.0$ Myr and then
rapidly declines to be 
$0.001 {\rm M}_{\odot}$ yr$^{-1}$) at $T=6$ Myr.

About 86\% of the gas can be converted into new stars within 13.6 Myr
in this model. This very high star formation efficiency 
(${\epsilon}_{\rm sf}$)
results from the fact that almost all of
the AGB ejecta can be accumulated very quickly
(in less than a few $10^6$ yr) within the central
region of the MSC owing to the deep gravitational potential
and consequently converted into new stars there.
The accumulated gas is  strongly self-gravitating  
(i.e., gas mass  comparable to stellar mass there) in the central
region of the cluster
so that the gas can continue to collapse to form new stars.
This result clearly suggests that {\it secondary star formation
within MSCs} is responsible for the origin of the observed
high stellar densities of the present GCs: 
A GMC can not be converted into a  GC just by one star-formation event.

A new  compact  star cluster embedded in low-density residual gas
can be formed in the central regions of the MSC within 13.6 Myr in this model.
Owing to the high star formation efficiency ($>0.5$),
the new cluster is highly likely to survive from gas removal by energetic
winds of massive stars and supernova explosion. 
As shown in Fig. 8,
the new cluster is much more compact than the original MSC and has 
a half-mass radius ($\sim 3$pc) significantly smaller than that of the
original MSC ($\sim 11$pc).
Owing to its compactness, the new cluster is much less susceptible
to tidal destruction by its host galaxy.
Thus a ``nested stellar system'' composed of a diffuse original cluster
(FG stars) and a compact new one (SG ones) can be formed as a result
of secondary star formation  from  AGB ejecta  within the MSC in this model.
This nested structure is one of common features of the simulated MSCs
in the present study.

Fig.9 shows how the fractions ($F_{\rm ns}$)
of the total mass of new stars ($M_{\rm ns}$) 
to the initial total mass of the MSC   ($M_{\rm s}$)
depend on $M_{\rm s}$ for $R<10$pc and $R<50$pc 
in the models with ${\rho}_{\rm th}=10^4$ atoms cm$^{-3}$.
Clearly there is a threshold cluster mass ($M_{\rm th}$) above 
which secondary star formation is possible: $M_{\rm th} \sim
6 \times 10^5 {\rm M}_{\odot}$ and does not depend on
${\rho}_{\rm th}$ in the present study.
Furthermore, $F_{\rm ns}$ is higher in MSCs with larger
$M_{\rm s}$, which implies that ${\epsilon}_{\rm sf}$ is
also higher in MSCs with larger $M_{\rm s}$.
Almost all stars can be formed within half-mass radii ($\sim 10$pc)
of MSCs with $M_{\rm s}>M_{\rm th}$ so that there can be
no differences  between $F_{\rm ns}$ for $R<10$pc and $R<50$pc.

\begin{figure}
\psfig{file=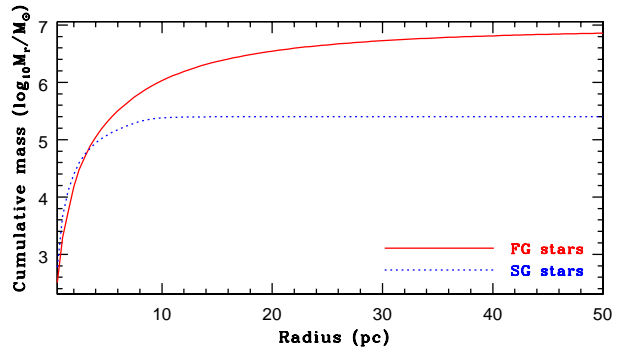,width=8.0cm}
\caption{
The same as Fig.11 but for the model with the more massive one
shown in Fig. 13.
}
\label{Figure. 14}
\end{figure}

\begin{figure}
\psfig{file=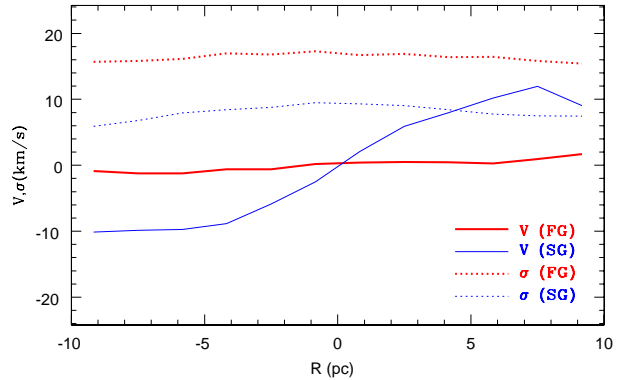,width=8.0cm}
\caption{
The radial profiles of rotational velocity $V$ (solid)  and one-dimensional
velocity dispersion $\sigma$ (dotted) for FG stars (red) and SG ones
(blue) in the more massive model shown in Fig. 13. Here $R$ is the
projected distance along the $x$-axis and the $y$-component of 
velocity for each star is used for estimating the line-of-sight velocity
and  velocity dispersion. These FG and SG stars correspond to old and new
stars, respectively, in the present study.
}
\label{Figure. 15}
\end{figure}

Although ${\epsilon}_{\rm sf}$ is rather high (0.86) in the model
shown in Figs. 7 and 8,  the total mass of new stars can be 
at most $4 \times 10^4 {\rm M}_{\odot}$ owing to the adopted
$f_{\rm AGB}=0.04$ that is reasonable for the standard IMF.
This means that MSCs with $M_{\rm s} \sim 10^6 {\rm M}_{\odot}$
can become low-mass GCs with multiple stellar populations.
Furthermore, ${\epsilon}_{\rm sf}$ is 0.3 for the model
with $M_{\rm s}=6 \times 10^5 {\rm M}_{\odot}$, which
suggests that the new cluster can significantly expand
after gas expulsion due to energetic winds of massive stars
and supernova.   The cluster is likely to become a low-mass,
low-density GC with a smaller fraction of SG stars
(like Pal 5). The dependence of ${\epsilon}_{\rm sf}$ on $M_{\rm s}$
implies that more massive GCs have larger fractions of SG stars.

The details of star formation histories depend weakly on 
$s_{\rm rot}$ and ${\rho}_{\rm th}$ such that
gas can be more rapidly converted into new stars in MSCs
with smaller $s_{\rm rot}$ and ${\rho}_{\rm th}$.
Fig. 10 shows that overall star formation rate is higher
in the model with $s_{\rm rot}=0$ 
(${\epsilon}_{\rm sf}=0.86$) than in that
with $s_{\rm rot}=0.08$ (${\epsilon}_{\rm sf}=0.47$).
This is mainly because initial angular momentum of the MSC
in the model with $s_{\rm rot}=0.08$ can suppress the formation
of a very compact gaseous region in the center of the MSC.
The secondary peak of the star formation rate 
in the model with $s_{\rm rot}=0.08$ at $T=2.8$ Myr
is due to later infall of gas  onto the MSC's center where a compact gas disk
is formed.

\begin{figure}
\psfig{file=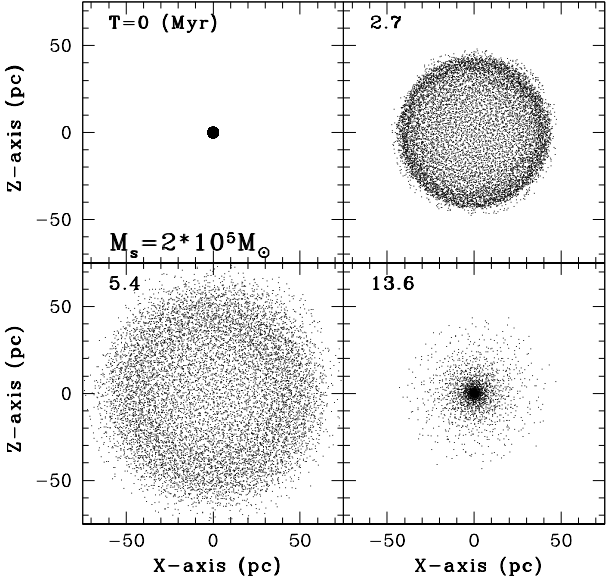,width=8.0cm}
\caption{
The same as Fig. 2 but for the model with the external gravitational 
field of the host galaxy of a MSC 
with $M_{\rm s}=2 \times 10^5 {\rm M}_{\odot}$.
The MSC is initially located in the center of its host.
}
\label{Figure. 16}
\end{figure}

\begin{figure}
\psfig{file=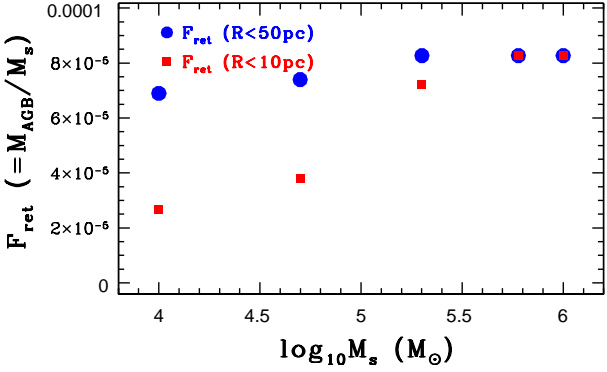,width=8.0cm}
\caption{
The same as Fig. 6 but for the models with
external gravitational fields of hosts. 
The MSCs are initially located in the centers of their  hosts.
}
\label{Figure. 17}
\end{figure}

\begin{figure}
\psfig{file=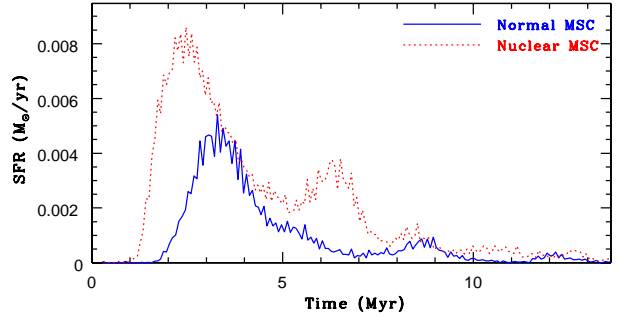,width=8.0cm}
\caption{
Star formation histories of low-density MSCs 
$M_{\rm s}=10^6 {\rm M}_{\odot}$,
$R_{\rm s}=75$pc,
$s_{\rm rot}=0.0$,
and
${\rho}_{\rm th}=10^4$ atoms cm$^{-3}$ in
two models
with (blue solid) and without (red dotted)
external gravitational fields of hosts. 
Normal and nuclear MSCs corresponds to models with
and without 
external gravitational fields of hosts, respectively.
}
\label{Figure. 18}
\end{figure}

\subsection{Structures and kinematics}

The formation of nested structures with very compact new clusters
and initially more diffuse old ones is one of essential ingredients
of the present simulations. Fig. 11, which shows a typical  example
of the nested structures, describes how the total stellar mass 
within $R$ depend on $R$ for old and new stars in the standard model.
Clearly the final cluster is dominated by new stars (SG) within $R<2$pc,
and the faction ($F_{\rm SG}$) of the total mass of new stars ($M_{\rm SG}$)
to the total mass of old stars ($M_{\rm FG}$)   and new ones
is more than 0.5 for $R<2$pc. Fig. 12 shows how $F_{\rm SG}$ depends
on $R$ for different models with different ${\rho}_{\rm th}$ and
$s_{\rm rot}$. Irrespective of these model parameters,
$F_{\rm SG}$ is higher in the inner regions of MSCs and lower in
the outer ones.  $F_{\rm SG}$ can be higher for higher ${\rho}_{\rm th}$
and smaller $s_{\rm sot}$.

As discussed in previous sections,  $M_{\rm s}$ should be at least
$2.5 \times 10^6 {\rm M}_{\odot}$ to explain GCs with 
$M_{\rm SG} \sim 10^5 {\rm M}_{\odot}$ (for $f_{\rm AGB}=0.04$).
Given that ${\epsilon}_{\rm sf}$ is not 1 ($0.3-0.9$ for most models)
and later dynamical evolution can reduce the total masses of MSCs,
stellar structures and kinematics of the simulated MSCs with rather
large $M_{\rm s}$ ($>5 \times 10^6 {\rm M}_{\odot}$)
can correspond to those of the observed ones of
GCs with masses larger than $\sim 10^5 {\rm M}_{\odot}$.
Fig. 13 shows the final spatial distributions of old and new stars
in the model with
$M_{\rm s}=10^7 {\rm M}_{\odot}$,
$R_{\rm s}=100$pc, 
$s_{\rm rot}=0.04$,
and 
${\rho}_{\rm th}=10^4$ atoms cm$^{-3}$.
Clearly the compact new cluster in the center of the MSC
is significantly flattened owing to gaseous dissipation during
gas accretion onto the central region.
This significantly flattened shape can be seen in new stars for most models
in which  $s_{\rm rot}$ is not 0 (i.e., even the models with $s_{\rm rot}=0.003$
show such flattened shapes).

The total mass of new stars ($M_{\rm SG}$) 
is larger than $10^5 {\rm M}_{\odot}$
and the effective radius ($R_{h,SG}$)
of the stars is 5.2pc, which is about
a factor of 5 smaller than that of the old ones ($R_{\rm h, FG}$).
The half-mass ratios ($=R_{\rm h,SG}/R_{\rm h, FG}$) are very similar
($\sim 0.2$) in the present models with different model parameters.
As shown in Fig. 14,  the new stars dominate the inner region ($R<$ 2pc)
of MSC, though $F_{\rm SG}$ there can not become so high as that
in the standard model. It is confirmed that more massive models 
with $M_{\rm s} = 10^7 {\rm M}_{\odot}$ 
with different $s_{\rm rot}$ and 
${\rho}_{\rm th}$ also show distinct  nested structures.

Fig. 15 shows clear differences in stellar kinematics between
old and new stars (thus FG and SG ones, respectively) in
the final MSC for the model. 
New stars clearly have rotational kinematics with the maximum
rotational velocity ($V_{\rm rot}$) of 12.0 km s$^{-1}$ 
and the smaller central velocity dispersion (${\sigma}_0$)
of 9.5 km s$^{-1}$, which means $V_{\rm rot}/{\sigma}_0 = 1.25$.
This rotational kinematics can not be seen in old stars,
which have $V_{\rm rot}/{\sigma}_0 = 0.1$ in this model.
The models with larger $s_{\rm rot}$ show 
larger $V_{\rm rot}/{\sigma}_0$ in new stars, which has already been
discussed  by Bekki (2010).
These results suggest that the present GCs originate from MSCs 
composed of two stellar populations with significantly different
kinematics.

\subsection{Influences of host galaxies}

Fig. 16 shows the time evolution of the projected  spatial distribution
of AGB ejecta for a nuclear MSC in the model without star formation
yet with the external gravitational  field of its host galaxy dominated
by dark matter halo ($M_{\rm h}=10^9 {\rm M}_{\odot}$).
In this model, only one  gaseous sphere of a AGB star
is included and located initially in the center of a nuclear  MSC with
$M_{\rm s}= 2\times 10^5 {\rm M}_{\odot}$
($s_{\rm rot}=0.0$) so that the results can be compared
with the model with $M_{\rm s}=2\times 10^5 {\rm M}_{\odot}$
shown in Fig. 2: It is much better to use these two comparative
models with only one gaseous sphere of a AGB star, because 
the effects of MSC's host galaxy can be more clearly demonstrated.
As shown in Fig. 16,
although the gas sphere expands initially to a larger radius ($R>50$pc),
the gas can finally return back to the central region of the MSC 
within $\sim 10$ Myr owing
to the deep potential well of its host galaxy.

Fig. 17 shows that $F_{\rm ret}$ in the models with external gravitational 
fields
of hosts do not depend so strongly on $M_{\rm s}$ as those without
the fields do: Irrespective of $M_{\rm s}$, most of AGB ejecta
can be retained within the central 50pc of MSCs 
in the models with external  gravitational fields of their 
host owing to their  much deeper gravitational potential wells.
However, $F_{\rm ret}$ within $R=10$pc depends more strongly
on $M_{\rm s}$ such that it is larger for larger $M_{\rm s}$.
This results indicates that inner gas densities ($R<10$pc)
can not be so high in nuclear MSCs with lower $M_{\rm s}$
so that
star formation can not occur efficiently in the MSCs.
It is actually confirmed that star formation can not occur efficiently
in the low-mass 
models with $M_{\rm s} \le 2 \times 10^5 {\rm M}_{\odot}$,
even if external gravitational fields are included.
Thus the  original masses of nuclear MSCs are still important
for secondary star formation processes  even if
they are located in nuclear regions of  their host galaxies.

External gravitational fields 
can enhance star formation of nuclear MSCs, in particular, for 
those with larger $M_{\rm s}$ and  lower 
mass-densities.
This enhancement can be most clearly seen in
Fig. 18 which describes
the time evolution of star formation rates 
for two models with 
$M_{\rm s}=10^6 {\rm M}_{\odot}$,
$R_{\rm s}=75$pc, 
$s_{\rm rot}=0.0$,
${\rho}_{\rm th}=0$ atoms cm$^{-3}$,
and with and without the external gravitational  field from a host with 
$M_{\rm h}=5 \times 10^9 {\rm M}_{\odot}$.
Star formation can start earlier and proceed more efficiently
in the model with the external gravitational field
(${\epsilon}_{\rm sf}=0.78$) than in the model without
(${\epsilon}_{\rm sf}=0.34$).
These  differences between two models with and without
external gravitational fields are less remarkable 
for high-mass and high-density MSCs, where star formation
can proceed efficiently even in isolation.

\begin{figure}
\psfig{file=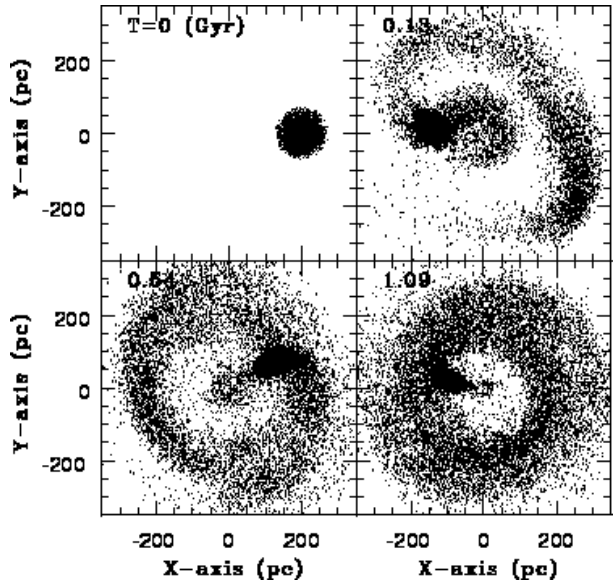,width=8.0cm}
\caption{
Time evolution of the distribution of stars  
projected onto the $x$-$y$ plane
for models with 
$M_{\rm s}=10^6 {\rm M}_{\odot}$,
$R_{\rm s}=75$pc,
$f_{\rm \sigma}=1.2$,
$M_{\rm h}= 5\times 10^9 {\rm M}_{\odot}$,
$r_{\rm s}= 1.6$kpc,
$x_{\rm s}=200$pc,
and
$f_{\rm v}=0.75$.
Time $T$ (in units of Gyr) that has elapsed since the simulation started
is shown in the upper left corner of each 
frame.
}
\label{Figure. 19}
\end{figure}

\begin{figure}
\psfig{file=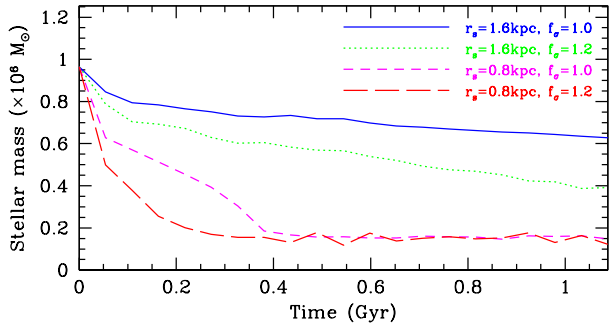,width=8.0cm}
\caption{
Total stellar masses within 100pc as a function of time $T$ 
for models with
$r_{\rm s}=1.6$kpc and $t_{\sigma}=1.0$ 
(blue solid),
$r_{\rm s}=1.6$kpc and $t_{\sigma}=1.2$ 
(green dotted),
$r_{\rm s}=0.8$kpc and $t_{\sigma}=1.0$ 
(magenta short-dashed),
$r_{\rm s}=0.8$kpc and $t_{\sigma}=1.2$ 
(red long-dashed).
For these for models,
$M_{\rm s}=10^6 {\rm M}_{\odot}$
$M_{\rm h}=5 \times 10^9 {\rm M}_{\odot}$
and  model parameters other than 
$r_{\rm s}$ and $t_{\sigma}$  are 
fixed. 
}
\label{Figure. 20}
\end{figure}

\begin{figure}
\psfig{file=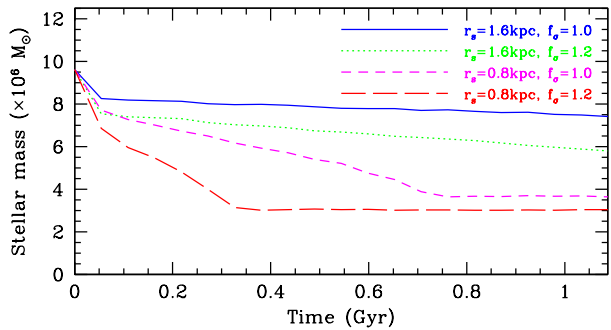,width=8.0cm}
\caption{
The same as Fig. 20 but for
$M_{\rm s}=10^7 {\rm M}_{\odot}$.
}
\label{Figure. 21}
\end{figure}

\section{Survival of originally massive stellar systems}

In the present scenario,  original MSCs need to keep substantial
masses at least for $\sim 10^8$ yr after they form within GMCs
so that gaseous ejecta from more massive AGB stars can be retained
in MSCs and finally converted into new stars (=SG ones).
The hosts (=massive dwarfs) can strip stars from MSCs owing to
their  tidal fields,  even if the fields are not so strong as
that of the Galaxy.
Therefore we here investigate how long MSCs can keep their substantial
fractions of stars when they are under strong gravitational influences
of their hosts. 

\subsection{The model}

We consider how a MSC dynamically evolves 
after most of residual gas left behind from the formation
of FG stars are expelled from the cluster through supernova
events.
Accordingly we adopt an assumption
that (i) the MSC is assumed to consists  purely of stellar particles 
and (ii) it is not necessarily in  virial equilibrium owing to
loss of the original mass of the GMC.
The cluster with a mass $M_{\rm s}$
and a size $R_{\rm s}$  has a Plummer  density profile 
with the scale length $a_{\rm s}$ ($R_{\rm s}=5a_{\rm s}$).
Owing to the above (ii),
the central velocity dispersion ${\sigma}_{\rm s}$ of the MSC  can be 
significantly larger than the one 
(${\sigma}_{\rm vir}$) which the MSC can have
when it is in  virial equilibrium.
Therefore we introduce  a parameter $f_{\sigma}$ that  is
the ratio of ${\sigma}_{\rm s}$ to ${\sigma}_{\rm vir}$ 
and ranges from 1.0 (virial equilibrium)
to  1.3. 
Thus MSCs can expand significantly for models with $f_{\sigma}>1$.

The cluster is influenced by the host with the density distribution
of the NFW halo (i.e., equation  (7)). 
The initial location of the MSC
with respect to the center of the host
is set to be
($x_{\rm s}$, $y_{\rm s}$, $z_{\rm s}$). 
The initial velocities of the MSC
in the $x$-, $y$-, and $z$-directions
are set to be
($u_{\rm s}$, $v_{\rm s}$, $w_{\rm s}$).
The orbital plane of the MSC is coincident with
the $x$-$y$ plane for all models in the present study
(i.e., $z_{\rm s}=0$ and $w_{\rm s}=0$).
The initial direction of the velocity vector of the MSC 
is only a parameter for the orbit of the MSC
owing to the spherically symmetric
distributions of the host and the MSC. 
We therefore assume that the initial direction
is parallel to the $y$-axis toward the positive 
$y$ (i.e., $v_{\rm s}>0$ and $u_{\rm s}=0$).
We assume $y_{\rm s}=0$, and thus 
$x_{\rm s}$ and $v_{\rm s}$ are free parameters
that determine the orbit of the MSC.

We consider that
the orbit of a  MSC is not necessarily circular within its host
and therefore introduce a free parameter $f_{\rm v}$ that
is the ratio of $u_{\rm s}$ to the circular velocity $v_{\rm cir}$
at $x_{\rm s}$ and ranges from 0.5 to 1.0 (i.e., circular orbit).
We mainly investigate two representative cluster models with 
$M_{\rm s}=10^6 {\rm M}_{\odot}$ and $R_{\rm s}=75$pc
and with
$M_{\rm s}=10^7 {\rm M}_{\odot}$ and $R_{\rm s}=100$pc
for different model parameters $M_{\rm h}$, $f_{\sigma}$, $x_{\rm s}$,
and $f_{\rm v}$.
Although we run numerous models, we mainly
describe the results of the models
with $M_{\rm h}=5 \times 10^9 {\rm M}_{\odot}$ and $r_{\rm s}=1.6$ kpc
($c=13.9$).

For each model,
we count the total number  of stars
that are located  within 100pc from the center of the MSC at each time
step in order to investigate the time evolution of the total stellar
mass of the MSC. Stars that are once stripped 
and by accident located  within the central 100pc 
without being bound by the MSC
can be  counted as those in the MSC.
We consider that
since the number fraction of these stars is very small,
these stars can only slightly 
overestimate the total stellar mass of  a MSC.
The initial stellar mass for a MSC is $(1-f_{\rm ej}) M_{\rm s}$
so that we can estimate the time evolution of the {\it stellar mass}
for the MSC.

\subsection{Results}

Fig.19 shows that the host of a MSC
can gradually remove stars from the outer part of the MSC
owing to the strong tidal field in the model with
$M_{\rm s}=10^6 {\rm M}_{\odot}$,
$R_{\rm s}=75$pc, 
$f_{\rm \sigma}=1.2$,
$M_{\rm h}= 5\times 10^9 {\rm M}_{\odot}$,
$r_{\rm s}= 1.6$kpc,
$x_{\rm s}=200$pc,
and 
$f_{\rm v}=0.75$.
About 60\% of original stars can be stripped 
within $\sim 1$ Gyr
and the stripped stars are 
distributed widely within the original orbital plane 
and thus regarded as ``field stars''.
Although the mass and the size of the MSC can become much
smaller than the original ones,  the MSC is still strongly
bound at $T=1.09$ Gyr.
The stripping process is gradual in this model: only 25\% of the original
mass of the MSC can be stripped within $0.1$ Gyr.
Thus, AGB ejecta might well be 
retained with the MSC to finally converted into
new stars (=SG ones) in this model.

Fig. 20 shows how the time evolution 
of total stellar masses within 100pc of MSCs  depends on $f_{\sigma}$
and $r_{\rm s}$ for models with
$M_{\rm s}=10^6 {\rm M}_{\odot}$,
$R_{\rm s}=75$pc, 
$M_{\rm h}= 5\times 10^9 {\rm M}_{\odot}$,
$x_{\rm s}=200$pc,
and 
$f_{\rm v}=0.75$.
It is clear from this figure that
stars can be more rapidly  stripped from MSCs in the models
with larger  $f_{\sigma}$ (=1.2)  and those in more compact hosts
($r_{\rm s}=0.8$kpc). The MSC in the model with $f_{\sigma}=1.2$
and $r_{\rm s}=0.8$kpc can lose $\sim 60$\% of its original mass
within $\sim 0.1$ Gyr owing to the combined effect of the more rapidly 
expanding stellar system and the stronger tidal field of the host.
In this mode,  AGB ejecta would not be so efficiently
retained and consequently  converted into new stars
within the MSC. The mass fraction of stripped stars ($F_{\rm strip}$)
is larger for models with smaller $r_{\rm s}$ and larger $f_{\sigma}$.

Fig. 21 shows how the time evolution 
of total stellar masses within 100pc of MSCs  depends on $f_{\sigma}$
and $r_{\rm s}$ for models with
$M_{\rm s}=10^7 {\rm M}_{\odot}$,
$R_{\rm s}=100$pc, 
$M_{\rm h}= 5\times 10^9 {\rm M}_{\odot}$,
$x_{\rm s}=200$pc,
and 
$f_{\rm v}=0.75$.
The results shown in Figs 20 and 21
indicate that 
stars can be more slowly and less efficiently
removed from MSC with larger $M_{\rm s}$
owing to their stronger gravitational potential wells.
MSCs with 
$M_{\rm s}=10^7 {\rm M}_{\odot}$ can not lose more than 50\% of
their original stellar masses within $\sim 1$ Gyr so that
AGB ejecta can be retained within the MSCs and converted into new stars.
These results imply that original masses of the present GCs with multiple
stellar populations should be rather large to keep substantial
fractions of their original stellar masses to be immune from tidal destruction
by their host within $\sim 0.1$ Gyr.

Initial densities of MSCs and $x_{\rm s}$ also can determine $F_{\rm strip}$
for a given set of parameters $M_{\rm s}$, $f_{\sigma}$, $M_{\rm h}$,
$r_{\rm s}$, and $f_{\rm v}$.
MSCs with lower densities and smaller $x_{\rm s}$ are more susceptible
to tidal destruction of their hosts so that they can lose more stars
within 1 Gyr (i.e., larger $F_{\rm strip}$). 
Thus the time evolution of stellar masses  of MSCs depends strongly on
a number of parameters (e.g., $M_{\rm s}$ and $f_{\sigma}$)
so that the time scale within which gaseous ejecta from AGB stars
can be retained and converted into new stars can also depend on
the parameters.
The present parameter study, however, indicates  that original MSCs are unlikely
to be destroyed completely within a timescale of $\sim 10^8$ yr for
most models: most MSCs can slowly ($>1$ Gyr)
disintegrate owing to the strong tidal fields of their hosts.
We thus suggest  that secondary star formation 
from AGB ejecta within MSCs is highly
likely for most MSCs with $M_{\rm s}=10^6-10^7 {\rm M}_{\odot}$
and can last at least $\sim 10^8$ yr.

\begin{figure}
\psfig{file=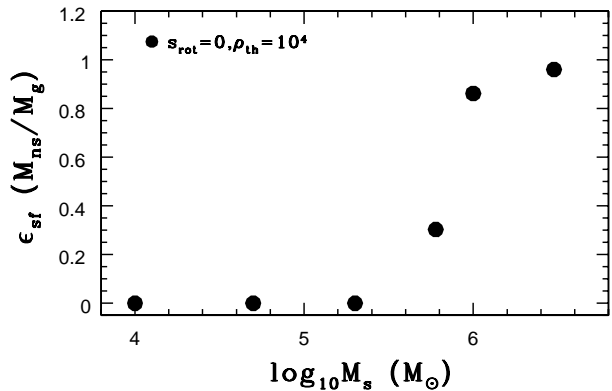,width=8.0cm}
\caption{
Dependences of star formation efficiencies ${\epsilon}_{\rm sf}$
($=M_{\rm ns}/M_{\rm g}$, where $M_{\rm g}$ is the 
initial total gas mass of a MSC) on $M_{\rm s}$ for 
$s_{\rm rot}=0$  and ${\rho}_{\rm th}=10^4$ atoms cm$^{-3}$.
%$s_{\rm rot}=0$ (blue filled squares) and $s_{\rm rot}=0.08$
%(red filled circles). 
}
\label{Figure. 22}
\end{figure}

\section{Discussion}

\subsection{A threshold cluster mass for secondary star formation}

We have shown that $M_{\rm s}$ can determine whether AGB ejecta from
FG stars of MSCs
can be retained within the  MSCs
and consequently converted into new stars efficiently.
Fig. 22 briefly summarizes  the present key results on how
$M_{\rm s}$ controles ${\epsilon}_{\rm sf}$
for the secondary star formation.
MSCs with larger $M_{\rm s}$ can have higher ${\epsilon}_{\rm sf}$
and 
we have confirmed that
his dependence does not depend on $s_{\rm rot}$ and ${\rho}_{\rm th}$.
MSCs with more rotation in FG stars are more likely to show
smaller ${\epsilon}_{\rm sf}$. 
Secondary star formation is possible if $M_{\rm s}$ is larger than
$6 \times 10^5 {\rm M}_{\odot}$, though ${\epsilon}_{\rm sf}$ is not
so high.
As long as $M_{\rm s} \ge 10^6 {\rm M}_{\odot}$,
${\epsilon}_{\rm sf}$ can be larger than 0.5 so that new compact
clusters, which would not disintegrate even after gas expulsion,
can form within old clusters.

Thus our results have shown  that there can be a threshold cluster mass 
($M_{\rm th} \sim [6-10] \times 10^5 {\rm M}_{\odot}$)
above which secondary star formation
with high ${\epsilon}_{\rm sf}$ can occur  within original clusters.
This ${\rm M}_{\rm th}$
can be as low as $\sim 2 \times 10^5 {\rm M}_{\odot}$,
if the original clusters are $\sim 10$ times  more compact than
the scaling-relations described by the equations (5) and (6) imply.
However ${\epsilon}_{\rm sf}$ is at most 0.1  even for such compact
clusters with lower masses.  On the other hand,
even if MSCs with $M_{\rm s} = 6 \times 10^5 {\rm M}_{\odot}$
are $\sim10$ times more diffuse than 
the scaling-relations described by the equations (5) and (6) imply,
their ${\epsilon}_{\rm sf}$ can be as large as 0.1. 
Therefore, it is reasonable to claim that $M_{\rm th}$ is 
$(6-10) \times 10^5 {\rm M}_{\odot}$.

D'Ercole et al. (2008) showed that secondary star formation within
original clusters is possible for $M_{\rm s}=10^7 {\rm M}_{\odot}$
but not for $M_{\rm s}=10^6 {\rm M}_{\odot}$. This implies
that $M_{\rm th}$ can be  somewhere between 
$10^6 {\rm M}_{\odot}$
and
$10^7 {\rm M}_{\odot}$
and is significantly larger than 
the derived $M_{\rm th}$ in the present study.
This is  mainly because modeling of
star formation and treatment of AGB ejecta are different between
these two works. Both models in D'Ercole et al. (2008) and the present
study have advantages and disadvantages  in describing evolution of AGB
ejecta within clusters. For example,  the present model enables
us to investigate self-consistently 3D hydrodynamical  interaction
of numerous gaseous spheres from AGB stars whereas the model
by D'Ercole investigated cooling of initially warm AGB ejecta
self-consistently.  More sophisticated models in our future papers 
need to investigate $M_{\rm th}$ in a fully self-consistent manner
by including physical effects that are treated 
by the present model in a simple  way.

\subsection{Star-to-star abundance variation dependent on cluster masses}

We here discuss how $M_{\rm s}$ determines 
nature of multiple stellar populations in GCs
owing to existence of $M_{\rm th}$.
We propose that $M_{\rm s}$ is one of key parameters which
can determine differences in ages and various chemical abundances
of light and heavy ones between multiple stellar populations. 
Based on the present numerical results,  
we divide MSCs into the following categories according to their $M_{\rm s}$.
It should be stressed here that $M_{\rm s}$ for a cluster
is not the present mass ($M_{\rm gc}$)
but its original one.

\subsubsection{$M_{\rm s}<10^4 {\rm M}_{\odot}$}

These low-mass clusters can neither retain AGB ejecta nor convert gas
into new stars owing to their shallow gravitational potential wells.
Therefore these clusters are highly unlikely to show
abundance spread even in light elements. 
Observations showed that the Galactic open cluster (OCs) 
show  unimodal distributions of CN band strength (Martell \& Smith 2009)
and do not have abundance spread
in light elements (e.g., Carretta et al. 2007)
and O-Na anti-correlation  (De Silva et al. 2009).
These low-mass clusters can be easily destroyed by interaction
with GMCs (e.g., Gieles et al. 2006)
so that they can not obtain fresh cold gas for further
star formation. 
The observed low-mass OCs in the Galaxy and the Magellanic Clouds
belong to  this class.
 
\subsubsection{$10^4 {\rm M}_{\odot} \le M_{\rm s}< 2 \times 10^5 {\rm M}_{\odot}$}

These clusters can not retain AGB ejecta efficiently within their central
regions so that secondary star formation can not occur. However, 
they can obtain fresh cold gas for further star formation by capture
and accretion  of cold molecular gas that are either those  left behind from
previous star formation events or those located close to them
(e.g., Bekki \& Mackey 2009). Therefore, if secondary star formation
occurs within these clusters, they can only show multiple stellar
populations with age differences yet no/little abundance spread even
in light elements.  The intermediate-age clusters in the LMC
and the SMC belong to this class.

\subsubsection{$2\times 10^5 {\rm M}_{\odot} \le M_{\rm s}<6\times 10^5 {\rm M}_{\odot}$}

Clusters with this mass range can retain most of their AGB ejecta  
within the central regions but can not convert the stars so efficiently
(${\epsilon}_{\rm sf}<0.3$)
into new stars within a short time scale ($<10^8$) yr.
Although dynamical fate of the accumulated AGB ejecta for the cluster remains
unclear,  ``triggering mechanisms'' such as  
cluster mergers (or initially very high-density clusters)
would be necessary to convert the gas so efficiently into new stars.
Strong ram pressure stripping of the gas by ISM of their host galaxies
could remove the gas from the clusters.

\subsubsection{$6 \times 10^5 {\rm M}_{\odot} \le M_{\rm s} < 10^7 {\rm M}_{\odot}$}

The ordinary Galactic GCs with abundance spread in light elements
can originate from clusters with this mass range, because
AGB ejecta can be retained in the clusters and converted into new stars
very efficiently
(${\epsilon}_{\rm sf}>0.5$).
The mass fraction of SG stars in a present GC depends mainly on
$M_{\rm s}$ and dynamical evolution of its original cluster
(i.e., destruction by its host and
long-term evolution driven by two-body relaxation).
If these massive clusters are initially stellar nuclei of 
massive dwarf galaxies, then AGB ejecta from SG stars
can be converted into
new stars (i.e., the formation of third generation of stars).  
This sequential formation of stars in the nuclear regions of hosts
can continue until the hosts are destroyed by much larger galaxies.

\subsubsection{$10^7 {\rm M}_{\odot} \le M_{\rm s}$}

Very massive clusters with this mass range are progenitors of
the present  massive GCs with
masses 
larger than  $10^6 {\rm M}_{\odot}$. Stellar systems with
$M_{\rm s}>10^7 {\rm M}_{\odot}$ would be dwarf galaxies themselves
rather than clusters, and therefore 
at least some fraction of massive GCs (e.g., $\omega$ Cen and G1)
may well the remnants of nucleated dwarf galaxies. 
Owing to much deeper gravitational potential wells of nucleated dwarfs,
not only AGB ejecta but also gas from more energetic massive stars 
and supernova would be able to be retained and finally
converted into new stars. Thus such massive GCs can show abundance spread
in both light and  heavy elements.

\subsection{Mixing of AGB ejecta  with fresh cold gas}

\subsubsection{Bondi accretion}

Using analytical models,
Pflamm-Altenburg \& Kroupa (2009) recently have shown that
if the mass of a cluster exceeds a threshold mass ($=10^6 {\rm M}_{\odot}$),
then the cluster can obtain a significant amount of warm ISM from outside
the cluster by Bondi accretion within a reasonable time scale
(i.e., a few Gyr).
We have confirmed their results using 3D hydrodynamical simulations,
though the threshold mass for gas accretion is larger 
($=3 \times 10^6 {\rm M}_{\odot}$) in our estimation
(Bekki et al. 2010).
We suggest that although secondary formation from 
mixed gas of AGB ejecta and the accumulated ISM
by Bondi accretion would be reasonable for massive GCs 
with possible age differences (like $\omega$ Cen),
such star formation has the following potentially serious
problems in explaining
chemical abundances of {\it ordinary} Galactic GC.

Firstly, if original clusters are within galaxies, then they can obtain
gas initially located in different regions with possible  different chemical
abundances  (owing to radial and azimuthal abundance gradients with the 
galaxies) thus are highly likely to have gas and new stars
 with different [Fe/H]. 
Therefore, only a minor fraction of the Galactic GCs may well have
SG stars formed from mixed gas of AGB ejecta and ISM obtained by Bondi 
accretion. Secondly, the Bondi accretion rate is too small 
($\sim 10^{-3} M_{\odot} $ yr$^{-1}$
 for $M_{\rm s}=3\times 10^6 {\rm M}_{\odot}$) for
typical ISM so that an enough amount of SG stars can not form
from gas chemically polluted only by massive AGB stars 
($> 5{\rm M}_{\odot}$) within an order of $10^8$ yr.
It is possible that if densities of warm ISM  are
as large as $10^2$ atoms cm$^{-3}$,
then clusters would be able to obtain an enough amount of gas
within a timescale of less than $10^8$ yr
(but if this is the case, then such high-density 
gas  may well be cold molecular clouds
rather than warm ISM).

Thirdly, even if warm ISM with initial temperature
of $\sim 10^4$K can be accreted by clusters,  further
efficient radiative cooling (to 10-100K) 
within a short timescale ($<10^8$ yr)  is required for
further star formation. Our previous and present study
have not yet investigated how warm ISM and AGB ejecta
mixed with and consequently convert into cold molecular
gas for further star formation  using a sophisticated model,
and accordingly we can not currently discuss the possibility
of secondary star formation from the accumulated warm ISM.
Our future more sophisticated models will  investigate whether
these three are really serious problems.

\subsubsection{Capture and accretion of cold molecular gas by clusters}

It is possible that the original MSCs would have a plenty of cold
molecular gas left behind from the formation of FG stars,
because the star formation efficiencies 
are highly likely to be well less than   100\%. Given that GMCs
which are progenitors  of FG stars should be very massive 
($>10^7 {\rm M}_{\odot}$) in the present scenario, then
the host GMCs would have numerous substructures
(i.e., smaller molecular clouds)  that remains intact
even after the formation of FG stars.
These molecular clouds can be captured  by or accreted onto 
MSCs, as a previous simulation demonstrated already 
(Bekki \& Mackey 2009). Owing to initially low temperatures of
the captured cold gas,  star formation from gas mixed between
AGB ejecta and the captured gas would proceed efficiently.

A potential problem  of this process is that it remain unclear whether
cold molecular clouds left behind the formation of FG stars can really
keep the same abundances as those of FG stars (i.e., remain ``pristine'')
without being chemically polluted by massive stars and supernova of 
FG stars. A significant fraction of the residual cold gas would be heated
up to become  high-temperature and low-density warm and hot gas 
so that they can not be captured
later by MSCs owing to their high-temperature and low-density.
Owing to clumpy structures of the original massive  GMCs,
a significant amount of energy from massive stars and supernova
can be expelled from low-density inter-substructure regions 
of the GMCs toward the halos of their hosts.
Some smaller cold molecular clouds that can survive from thermal
and kinetic feedback effects of energetic stars would be able to
be captured later by MSCs and consequently mixed with AGB ejecta.

The above processes of mixing of cold molecular gas and AGB ejecta
have not been investigated in detail by any previous 3D hydrodynamical
simulations of star-forming GMCs. Bekki \& Chiba (2007) investigated
(i) how massive stars and type II supernova influence gas dynamics
within star-forming GMCs and (ii) chemical abundances of gas left behind
from star formation. They however did not include ejection of gas
so that they could not investigate
how AGB ejecta mixed  with the residual gas to form new stars.
It is thus doubtlessly worthwhile for our future more sophisticated
models to investigate whether and what fraction of
residual molecular cloud left behind
from the formation of FG stars can be converted into new stars without
being chemically polluted by massive FG stars.

\subsection{Origin of  GCs with multiple stellar populations
 in the Magellanic Clouds}

A number of recent observational studies have found that
a significant fraction of intermediate-age cluster in
the LMC and the SMC show 
double main sequence turnoffs (hereafter referred to as DMSTOs
for convenience) on their color magnitude diagrams 
(Mackey \& Broby Nielsen 2007;
Mackey et al. 2008; Glatt et al. 2008;
Goudfrooij et al.
2009; Milone et al. 2009). 
The simplest explanation is that the observed DMSTOs in each of
these clusters represent two distinct stellar populations with
differences in age of $\sim 100 - 300$ Myr. 
Recently Bastian \& de Mink (2009) have proposed that
stellar rotation in stars with masses between 1.2 and 1.7${\rm M}_{\odot}$
can mimic the effects of DMSTOs without resorting to age differences
between stellar populations within clusters of the Magellanic Clouds.
However, the presence of a dual red clump of giant stars in
the color magnitude diagram of NGC 419 in the SMC  is
suggested to be inconsistent with the claim by Bastian \& de Mink (2009)
but be explained naturally by the presence of multiple star-formation
episodes (Rubele et al. 2010).

Mucciarelli et al. (2008) investigated
chemical abundances of light odd-Z, $\alpha$, iron-peak, and neutron-capture 
elements for four intermediate-age clusters in the LMC
and found negligible star-to-star scatter for them.
However, Mucciarelli et al. (2009) revealed significant abundance
inhomogeneities in [Na/Fe], [Al/Fe], [O/Fe], and [Mg/Fe] and
O-Na and Mg-Al anti-correlations in old GCs ($>10$ Gyr).
These observations imply that intermediate-age clusters
have (at least) two generations of stars with different ages
yet similar chemical abundances whereas old ones are
like the Galactic GCs with abundance spread in light elements.
Therefore, any theory of GC formation need to explain
why the LMC appears to have two different types of
GCs with multiple stellar populations.

Bekki \& Mackey (2009) showed that  low-mass clusters 
($M_{\rm s} \sim 5 \times 10^4 {\rm M}_{\odot}$)  
can capture cold  GMCs and thereby convert the gas into new stars
within their central regions in the LMC.
The present study has shown that such low-mass clusters can not
efficiently retain AGB ejecta and thus are unable to use
the ejecta for further star formation.
The results by Bekki \& Mackey (2009) and the present study
therefore combine to suggest that if the observed DMSTOs 
in clusters of the LMC are due to age differences of stellar populations,
then the two populations are highly unlikely to show abundance spread
in light elements owing to their low masses ($<10^5 {\rm M}_{\odot}$).

The present study also suggests
that the old clusters of the LMC can show abundance spread in light elements,
mainly because their original clusters are significantly more
massive than the present ones with masses 
ranging from  $10^5 {\rm M}_{\odot}$ to $6 \times 10^5 {\rm M}_{\odot}$.
Recently Conroy \& Spergel (2010) have suggested that clusters with
masses larger than a threshold mass of $\sim 10^4 {\rm M}_{\odot}$ can 
form SG stars from  AGB ejecta mixed with ISM accreted onto
clusters. 
The observed lack of abundance inhomogeneity
in light elements in the intermediate-age
clusters of the LMC (Mucciarelli et al. 2008) 
seems to be   inconsistent with their model.
The present study suggests that the threshold mass for the formation
of SG stars from  AGB ejecta mixed with  pristine gas 
is much larger than the one suggested by Conroy \& Spergel (2010). 

Old and metal-poor ([Fe/H]$<-1.65$) clusters in the LMC
have masses more than $10^5 {\rm M}_{\odot}$ 
(Mackey \& Gilmore 2003) and would have
lost a significant fraction of their masses due to long-term
internal evolution effects and external tidal field of the LMC.
Therefore recent observations on star-to-star abundance
variations in light elements in old and intermediate  clusters of the LMC
by Mucciarelli et al.  (2008, 2009) 
strongly suggest that the threshold cluster mass ($M_{\rm th}$) for secondary
star formation from AGB ejecta mixed with pristine gas
would be much more than $10^5 {\rm M}_{\odot}$.
These observations are consistent with the present model
with $M_{\rm th} \sim (6-10) \times 10^5 {\rm M}_{\odot}$  and the one
by D'Ercole et al. (2008) which showed  that clusters with
$M_{\rm s} \sim 10^7 {\rm M}_{\odot}$ can form SG stars from
AGB ejecta.

\subsection{Were original massive stellar systems  really ``clusters'' ?}

D'Ercole et al. (2008) first claimed that original single 
stellar systems (``clusters'') 
composed purely of FG stars should have masses of $\sim 10^7 {\rm M}_{\odot}$
to explain the observed masses of massive GCs ($\sim 10^6 {\rm M}_{\odot}$).
Although our present simulations confirm their claim,
it remain unclear how such massive single clusters  form from very
massive GMCs. Given the possible substructures  within host GMCs for
FG stars,  the original stellar systems
can be still clusters of smaller clusters when FG AGB stars
to eject their gaseous winds. If this is the case,
evolution of AGB ejecta in such clusters of clusters would be significantly
different from what D'Ercole et al. (2008) and the present study
describe.

It would be also possible that (i) original massive stellar systems of GCs 
are dwarf galaxies themselves and therefore (ii) compact
stellar systems formed from AGB stars of the dwarfs
are identical to
their stellar galactic nuclei. In this scenario,
stellar nuclei are dominated by SG stars that can be formed 
more quickly from
AGB ejecta of FG ones owing to much deeper gravitational potentials
of their host with dark matter halos.
Therefore,  stellar nuclei of nucleated dwarfs are highly likely to be dominated
by He-rich, Na-rich, and O-poor stars. 
Future observations on chemical abundances of stars in nuclei of
the Sagittarius dwarf and NGC 205 will enable us to discuss
whether this scenario  is physically viable.

\subsection{Implications}

\subsubsection{A bottom-heavy IMF for SG stars ?}

The present simulations have shown that secondary star formation
from AGB ejecta can occur at the central regions of original clusters,
where mean stellar number densities exceed $\sim 2 \times 10^2$ 
stars  pc$^{-3}$ 
(for $M_{\rm s}=10^6 {\rm M}_{\odot}$).
Recently Krumholz et al. (2009) investigated formation of massive 
stars by gaseous accretion within a gas cloud with a mass of
$100 {\rm M}_{\odot}$ and a size of 0.1 pc
and found that two (binary) stars with masses of $29.2 {\rm M}_{\odot}$
and $41.5 {\rm M}_{\odot}$ can be formed within $5.7 \times 10^4$ years.
Therefore the present simulations imply that
gas clouds for massive stars (as those investigated by Krumholz et al. 2009
above) can interact frequently with stars within 
the central regions of a dense stellar system
owing to
the small mean separation of stars ($\sim 0.2$ pc) and
the short dynamical time scale  ($\sim 5\times 10^5$ yr).
It would be accordingly possible that massive star formation is severely
suppressed owing  to this star-cloud interaction which can
prevent  gas accretion onto clouds forming massive stars:
IMFs for the formation of SG stars  
within MSCs would be bottom-heavy.
If formation of  binary star-forming gas clouds,
which are progenitors  of binary stars, can be also prevented owing
to star-cloud interaction, then the observed small 
binary fractions among SG stars
in GCs (D'Orazi et al. 2010b) 
can  be naturally explained.

If only SG stars have a bottom-heavy IMF with the power-law slope
of $3.35$ with FG ones with $\alpha = 2.35$, then
the number fraction of low-mass stars with masses lower than
$0.5 {\rm M}_{\odot}$ for SG stars to that for FG ones 
is 1.7 for $m_{\rm u}=120 {\rm M}_{\odot}$. This number fraction can become
as large as 2 by assuming unrealistically
 low $m_{\rm u}$ ($<1 {\rm M}_{\odot}$)
and steeper $\alpha$ ($<3$).
Therefore, if IMFs can be  bottom-heavy
only for SG stars, 
then the required original masses of clusters that finally become
the present GCs can be up to by a factor of 2
smaller than those derived
in the present work with the standard IMF both for FG and SG stars.
Thus, the possible bottom-heavy IMF can not change the present
main conclusion that the progenitor clusters of the present GCs
are much more massive than the present GCs.

\subsubsection{Origin of rotation, shapes, and scaling-relations  of GCs}

Previous works discussed 
the origin of shapes and rotation 
(e.g., Frenk \& Fall 1982; Einsel \& Spurzem 1999),
sizes and luminosities (e.g., Murray 2009),
scaling-relations (e.g., Djorgovski 1993; Bekki et al. 2004; 
Gieles et al. 2010) using analytical models and collisionless numerical
simulations.  They accordingly did not discuss the origin of GC properties
in the context of initially nested structures and ignored the importance
of gas dynamics  in the formation of  GC properties.
If most GCs really originate from nested stellar systems, then
conclusions of the above previous works based on models
with ``non-nested'' stellar
structures must be dramatically modified.
Thus it would be reasonable to say that the above conclusions
can apply only for low-mass clusters  with single stellar
population and no nested structures.

Bekki (2010) recently has shown that the origin of 
the observed rotation in GCs
(e.g., Meylan \& Mayor 1986; Meylan \& Heggie 1997; Anderson \& King 2003)
can be closely associated with dissipative gas dynamics of AGB ejecta
from FG stars within forming GCs.
Given the observed large mass fractions of SG stars in the present GCs
(e.g., Carretta et al. 2010),
gas dynamics of AGB ejecta within original clusters composed only of
FG stars may well be a key determinant for structure and kinematics
of the present GCs. 
Although recent numerical simulations have investigated long-term
dynamical evolution of GCs with initially nested stellar structures
(e.g., Decressin et al. 2008; D'Ercole et al. 2008),
their models do not include initial rotation and flattened shapes
of SG stars so that their results can not be used for discussing
the origin of rotation and shapes of GCs.
Our future long-term ($>10$ Gyr) dynamical simulations
of GCs  based on structures and kinematics of GCs
with nested stellar structures,  flattened 3D distributions,
and rotation simulated in the present study will enable
us to discuss structure and kinematics in a consistent manner.

\subsubsection{Where are forming GCs with nested structures ?}

Recently Vink\'o et al. (2009) have found two stellar populations 
with younger and older ones
with  ages of $10-16$ Myr and $32-100$ Myr, respectively,
in the young, massive stellar cluster Sandage-96 in a spiral
arm of NGC 2403. Given the observed possible large age difference
(up to $\sim 100$ Myr), this type of young massive clusters
can be the candidates  that are now forming or have just formed
SG stars from AGB ejecta from FG ones.
Vink\'o et al. (2009) also have found that the younger population 
are located closer to the center of the cluster in comparison with
the older one. 
If the cluster has a mass as large as $10^6 {\rm M}_{\odot}$,
the observation would be consistent with the present results on
the nested structure of MSCs, though quantitative comparison 
(i.e., the mass-ratio and the half-mass-radius-ratio of the two populations)
with the present results is not currently possible.
The present study suggests
that if the total mass of the cluster is as large as $10^6 {\rm M}_{\odot}$,
the redder colors in the younger population of the cluster
can be due to gas and dust that now surrounds
the younger population and were previous  ejected from the older one.

If future observations find young clusters with 
large masses ($>10^6 {\rm M}_{\odot}$) and two populations with 
the age differences as large as $10^8$ yr 
in actively star-forming galaxies,  
they can be forming GCs with nested structures and thus
provide strong constraints on the formation models of GCs like
the present one.
As suggested above,
new compact clusters can be shrouded by
gas and dust left behind from the formation of SG stars until
supernova events expel the gas and dust: the dust-shrouded new clusters
would have very red color for their young ages and possibly they
can not be even seen in optical bands if dust extinction is so heavy. 
If FG stars can not be clearly identified as compact  clusters
owing to their initially more diffuse distributions,
then dust-shrouded new clusters within the FG stars might well
be identified  as giant HII regions with no optical counterparts. 
However it is not clear whether the observed giant HII regions
with no near-infrared cluster counterparts in starbursting luminous
infrared galaxies with numerous super-star clusters
(e.g., Alonso-Herrero et al. 2002)  can be
dust-obscured very young compact  clusters
(SG)  embedded in older ones (FG).

Massive stellar systems with masses of $\sim 10^7 {\rm M}_{\odot}$
were discovered in ultra-luminous infrared-galaxies  with 
dust-shrouded strong
starbursts (e.g., Monreal-Ibero et al. 2007). Numerical simulations
showed that such dusty starburst galaxies can evove into ``E+A''
galaxies
with poststarburst populations with ages of $0.1-1$ Gyr
(Bekki et al. 2001).  The present  sutdy has shown that
MSCs with $M_{\rm s}=10^7 {\rm M}_{\odot}$ and ages of less than 1 Gyr
(before destruction of the old clusters) can have nested steller 
structurs.  These results combine to imply that
young GCs with nested structures are highly likely to exist in
E+A galaxies. It is thus worthwhile for future observational studies
to search for massive young clusters in  E+A galaxies 
(e.g., NGC 5102) to provide
a clue to the origin of multiple stellar populations of GCs.

%\subsection{A more realistic model}
%Merger of smaller subunits of smaller clusters 

\section{Conclusions}

We have performed 3D numerical simulations of star formation from AGB ejecta 
in  MSCs formed from GMCs within their host galactic building blocks.  
We have considered that
stars initially in MSCs and  those newly
formed from AGB ejecta corresponds to FG and SG stars,
respectively,  in the present
GCs and thus discussed the origin of multiple stellar populations of
GCs. 
The main results are summarized as follows.

(1) Gaseous ejecta from AGB stars can be retained within MSCs
with $M_{\rm s}$ larger than $\sim 2\times  10^5 {\rm M}_{\odot}$,
if MSCs are isolated (i.e., not being influenced by ram pressure of ISM). 
The mass fraction of gas retained in a MSC to $M_{\rm s}$
($F_{\rm ret}$) depends on $M_{\rm s}$ such that $F_{\rm ret}$ is larger
in larger $M_{\rm s}$.
The multiple hydrodynamical interaction of gaseous ejecta from
numerous AGB stars can significantly  increase gas mass accumulated
within the central regions of MSCs, which was first found by the 
present study.

(2) If $M_{\rm s}$ of MSCs exceed a threshold cluster mass $M_{\rm th}$,
gaseous ejecta of AGB stars can sink into the central regions
and then form high-density gaseous regions so that
new stars (i.e., SG stars) 
can be formed there with high star formation efficiencies ($>0.3$).
$M_{\rm th}$ is demonstrated  to be $(6-10) \times 10^5 {\rm M}_{\odot}$
in the present study.
Deep gravitational potentials of MSCs play a great role
both in retaining the accreting AGB ejecta onto the central regions
and in forming high-density gaseous regions there.
Therefore secondary star formation within clusters 
is inevitable in MSCs with $M_{\rm s}$ larger than $M_{\rm th}$.  
Owing to the existence of $M_{\rm th}$,  young open clusters and super star 
clusters with $M_{\rm s} < 
6 \times 10^5 {\rm M}_{\odot}$ can not evolve into stellar
systems with abundance spread in light elements.

(3) Owing to the high star formation efficiencies 
(${\epsilon}_{\rm sf}=0.3-0.9$), 
very compact stellar systems can be formed from AGB ejecta
within a time scale of
$\sim 10^7$ yr (after massive AGB stars start to eject their winds).
The half-mass radius of  a new compact cluster
is by a factor of 5 smaller than that of the old one for most 
models in the present study.
The nested clusters (or ``cluster within cluster'') 
with more diffuse old cluster and very compact new one
are characteristic
for the present numerical models.

(4) At most $1-4$\% of all stars can be from new stars formed
from AGB ejecta for the canonical IMFs in the present study.
Therefore, in order to explain the present total mass of SG stars 
with masses of $\sim 10^5 {\rm M}_{\odot}$ in a GC,
$M_{\rm s}$  is required to be  as large as 
$(3-10) \times 10^6 {\rm M}_{\odot}$.
The required mass for the progenitor MSC for $\omega$ Cen 
can be as large as $10^8 {\rm M}_{\odot}$ owing to its larger
present mass. 
Given that ${\epsilon}=0.3-0.9$ for secondary star formation,
the original masses of progenitors  ($M_{\rm s}$)
for the present GCs with masses of $M_{\rm gc}$
need to be larger than $25M_{\rm gc}$.

(5) Structural and kinematical properties of new compact
clusters formed within original old ones 
depend on $s_{\rm rot}$ such that new clusters can be more flattened
and more strongly supported by rotation in MSCs with larger $s_{\rm rot}$.
The new clusters are much more compact and more strongly supported by
rotation than the old ones in all models with initial rotation in 
old clusters. These structural and kinematical differences between
old and new clusters do not depend on model parameters
such as $M_{\rm s}$, $R_{\rm s}$,  and ${\rho}_{\rm th}$.
Given that the present GCs are dominated by SG stars,
the result implies that the physical origins of flattened shapes and
internal rotation of GCs are closely associated with formation processes
of SG stars through gas dynamics within MSCs.

(6) Evolution of MSCs initially in the very centers of their hosts
can be significantly different from that of MSCs outside the centers,
mainly because AGB ejecta can be more effectively retained within
nuclear MSCs owing to much deeper gravitational potentials of their hosts.
Irrespective of $M_{\rm s}$, most AGB ejecta can be retained in
these nuclear  MSCs.  Star formation, however, can proceed within
nuclear MSCs, only if $M_{\rm s}$ exceeds a threshold value that is 
almost the same as $M_{\rm th}$
described above.  
Strong gravitational fields of hosts can enhance secondary star formation
of nuclear MSCs.
Multiple episodes of star formation (i.e., third and fourth
generations of stars)  are  possible  
in nuclear MSCs with $M_{\rm s} \ge M_{\rm th}$
until their hosts are destroyed during accretion
of the hosts on larger galaxies.
The origin of multiple generations of stars observed in $\omega$ Cen and
NGC 2808 can be closely associated with evolution of  nuclear MSCs.

(7) MSCs can lose gradually ($\sim 0.1-1$ Gyr)   significant
fractions  of their stars during dynamical evolution
within their hosts owing to tidal stripping by their hosts.
The stripping processes 
and the timescale depend strongly on $M_{\rm s}$, orbits,
and  initial locations of MSCs within hosts.
If stars in a MSC
enter into their AGB phases after the MSC has lost most of their stars
because of tidal stripping,
then the AGB ejecta can not be efficiently retained in the 
MSC owing to its  much shallower gravitational potential.
Therefore gaseous ejecta from AGB stars with lower masses are less likely to
contribute to secondary star formation owing to the destruction
of the MSC by its host.
Thus the destruction of MSCs
due to their hosts can provide a mechanism by which star formation
can be truncated in MSCs.

(8) Direct capture and accretion of cold molecular clouds by MSCs themselves
can be much more efficient ways to dilute the AGB ejecta of MSCs
(i.e., mixing of these gaseous components)
in secondary star formation within MSCs. The molecular clouds need to
have chemical abundances very similar to those of MSCs, and thus they
are either
(i) those that are located very close to MSCs
or (ii) those left behind from the formation of FG stars.
The observed hierarchical structures of GMCs are suggested to be responsible
for the dilution processes during formation of SG stars.

(9) Low-mass MSCs with $M_{\rm s} < 10^5 {\rm M}_{\odot}$ in the LMC can obtain
cold molecular gas 
by capture of the gas but can not retain AGB ejecta
owing to much shallower gravitational potentials of MSCs.
Therefore the intermediate-age
clusters with DMSTOs recently discovered in the Magellanic Clouds
can have multiple stellar populations with age differences yet no
abundance spread in light elements owing to their low masses.
The initial masses of the old GCs in the LMC were  significantly
larger than their present masses of $(1-6) \times 10^{5} {\rm M}_{\odot}$
so that they could retain AGB ejecta to form  SG stars with chemical
abundances different from those of FG ones.

(10) It is suggested that formation of massive stars 
can be severely suppressed in very dense  central
regions of MSCs (i.e., a bottom-heavy IMF), because dynamical interaction between stars
and forming molecular cores (where individual stars can form)
can prevent the growth of the cores.  
If dense stellar systems  are really environments within which  only
low-mass stars can form (owing to a bottom-heavy IMF), then 
the original masses of the present 
GCs can be to some extent lower than
the proposed one ($M_{\rm s} > 25M_{\rm gc}$) 
in the present scenario. \\

Thus GCs with the observed high central stellar densities  
($\sim 10^4$ stars pc$^{-3}$)
can not be formed directly from GMCs 
with single star formation events within GMCs
in a scenario explored  in the present study.   The origin 
of the observed high stellar densities
of GCs  can
be associated with unique star formation processes within
central regions of  already compact stellar systems (composed of
FG stars). 
The deep gravitational potential wells of original clusters
with much lower stellar densities ($\sim 10^2$ stars pc$^{-3}$
at half-mass radii) can play a role in enhancing
dramatically star formation efficiencies and consequently
forming new compact clusters.

The progenitor stellar systems for the present GCs with
multiple stellar populations need to be very massive
($M_{\rm s} > 25 M_{\rm gc}$) to retain effectively
AGB ejecta from FG stars,  dilute the gas with cold molecular gas,
form new stars from the gas,
and  finally have the total stellar masses
of SG stars  similar to the observed
ones  in the Galactic  GCs. 
The present scenario explains why open clusters in the Galaxy
and intermediate-age GCs in the LMC do not show clear abundance
spread in light elements. If GCs really originate from MSCs
with large $M_{\rm s}$, then the host GMCs for such MSCs
should be even more massive ($10^7-10^8 {\rm M}_{\odot}$). 
We plan to investigate how very massive GMCs can form 
and evolve into MSCs due to efficient star formation
within  massive dwarf galaxies at high $z$.

\section{Acknowledgment}
I am   grateful to the  referee Raffaele Gratton for valuable comments,
which contribute to improve the present paper.
KB acknowledges the financial support of the Australian Research
Council
throughout the course of this work.
I am   grateful to all participants in the last conference
``Multiple stellar populations in globular clusters ''
held in Asiago (Italy)  on 14th  
$-$ 16th  Sep 2010 for their having extensive discussion on origin
of globular clusters 
with me,
which helped  me better interpret the present numerical results
in terms of the corresponding observational ones.
Numerical computations reported here were carried out on the GRAPE
system
at ICRAR in  the University of Western Australia  and that
kindly made available by the Center for computational
astrophysics
(CfCA) of the National Astronomical Observatory of Japan.

\end{document}